\documentclass[journal,10pt]{IEEEtran}
\usepackage{amssymb}
\usepackage{amsmath}
\usepackage{epsfig}
\usepackage{epstopdf}
\usepackage{cite}
\usepackage{bm}
\usepackage[noend]{algpseudocode}
\usepackage{lettrine}
\usepackage{algorithmicx,algorithm}
\usepackage{color}
\usepackage{subfigure}
\usepackage{balance}
\IEEEoverridecommandlockouts

\graphicspath{ {figures/}}

\begin{document}
	
	\title{MU-MIMO Symbol-Level Precoding for QAM Constellations with Maximum Likelihood Receivers}

	\author{Xiao Tong,~\IEEEmembership{Graduate~Student~Member,~IEEE}, Ang Li,~\IEEEmembership{Senior~Member,~IEEE}, Lei Lei,~\IEEEmembership{Senior~Member,~IEEE},\\ Xiaoyan Hu,~\IEEEmembership{Member,~IEEE}, Fuwang Dong,~\IEEEmembership{Member,~IEEE}, Symeon Chatzinotas,~\IEEEmembership{Fellow,~IEEE}\\ and Christos Masouros,~\IEEEmembership{Fellow,~IEEE}
	\thanks{
		
	X. Tong and X. Hu are with the School of Information and Communications Engineering, Faculty of Electronic and Information Engineering, Xi’an Jiaotong University, Xi’an, Shaanxi 710049, China (e-mail: xiao.tong.2023@stu.xjtu.edu.cn; xiaoyanhu@xjtu.edu.cn).
	
	A. Li and L. Lei are with the School of Information and Communications Engineering, Faculty of Electronic and Information Engineering, Xi’an Jiaotong University, Xi’an, Shaanxi 710049, China, and also with the National Mobile Communications Research Laboratory, Southeast University, Nanjing 210096, China (e-mail: ang.li.2020@xjtu.edu.cn; lei.lei@xjtu.edu.cn).
	
	F. Dong is with the College of Intelligent Systems Science and Engineering, Harbin Engineering University, Harbin 150001, China (e-mail: dongfuwang@hrbeu.edu.cn).
	
	S. Chatzinotas is with Interdisciplinary Center for Security, Reliability and Trust (SnT), University of Luxembourg, 4365 Esch-sur-Alzette, Luxembourg (e-mail: symeon.chatzinotas@uni.lu).
	
	C. Masouros is with the Department of Electronic and Electrical Engineering, University College London, London WC1E 7JE, U.K. (e-mail: c.masouros@ucl.ac.uk).
	}
	}

	\markboth{}
	{}
	
	\maketitle
	
	\begin{abstract}
	In this paper, we investigate symbol-level precoding (SLP) and efficient decoding techniques for downlink transmission, where we focus on scenarios where the base station (BS) transmits multiple QAM constellation streams to users equipped with multiple receive antennas. We begin by formulating a joint symbol-level transmit precoding and receive combining optimization problem. This coupled problem is addressed by employing the alternating optimization (AO) method, and closed-form solutions are derived by analyzing the obtained two subproblems. Furthermore, to address the dependence of the receive combining matrix on the transmit signals, we switch to maximum likelihood detection (MLD) method for decoding. Notably, we have demonstrated that the smallest singular value of the precoding matrix significantly impacts the performance of MLD method. Specifically, a lower value of the smallest singular value results in degraded detection performance. Additionally, we show that the traditional SLP matrix is rank-one, making it infeasible to directly apply MLD at the receiver end. To circumvent this limitation, we propose a novel symbol-level smallest singular value maximization problem, termed SSVMP, to enable SLP in systems where users employ the MLD decoding approach. Moreover, to reduce the number of variables to be optimized, we further derive a more generic semidefinite programming (SDP)-based optimization problem. Numerical results validate the effectiveness of our proposed schemes and demonstrate that they significantly outperform the traditional block diagonalization (BD)-based method.
	\end{abstract}
	
	% Note that keywords are not normally used for peerreview papers.  To address this coupled problem, we employ the alternating optimization (AO) method to obtain two subproblems that can be optimized alternately.

	\IEEEpeerreviewmaketitle
	
	\section{Introduction}\label{introduction}
	\lettrine[lines=2]{W}{ith} the worldwide deployment of fifth-generation (5G) mobile communication networks, both academia and industry have envisioned the roadmap to the future sixth-generation (6G) wireless communication system, which aims to achieve massive, hyper-reliable, and low-latency communications\cite{Tutorial MIMO,Sensing}. Multiple-input multiple-output (MIMO) communication is one of the promising technologies to meet these requirements by utilizing multiple antennas on both transmitter and receiver sides to simultaneously transmit multiple data streams to multiple users (MU)\cite{Tutorial Large-Scale MIMO}. This spatial multiplexing approach increases communication capacity and enhances spectral efficiency. However, a primary limitation of MIMO systems is the interference arising from the simultaneous transmission of multiple signals using the same time and frequency resources. Therefore, it is crucial for both the transmitter and receiver to exploit the additional degrees of freedom offered by multiple antennas to mitigate the interference\cite{ZF,RZF,BD,RBD,SLP Surveys,Selective-Precoding,DI,CF-PSK,CF-QAM,Group SLP,Large-scale,DACs,TCOM SER,WCL SER,Con JTRSLP,Jour JTRSLP,PSK SLP MU-MIMO} and improve decoding efficiency\cite{Fifty years mimo detection,LOW complxity MIMO Detection,ZF Detection,MMSE Detection,MAME Detection,MLD,QRMLD,QRMMLD}.
	
	\subsection{Previous Work on Transmit Precoding}\label{PWOTP}
	At the transmitter side, precoding has been extensively studied as an effective strategy to mitigate interference and enhance communication performance. Traditional precoding schemes utilize channel state information (CSI) to design the precoding matrix that maximizes some system performance while adhering to the power budget. Representative examples include zero-forcing (ZF)\cite{ZF}, regularized zero-forcing (RZF)\cite{RZF}, block diagonalization (BD)\cite{BD} as well as regularized block diagonalization (RBD)\cite{RBD}. However, these schemes primarily focus on interference suppression, overlooking the potential benefits of instantaneous interference, which can be further exploited at the symbol level, a technique known as symbol-level precoding (SLP)\cite{SLP Surveys1, SLP Surveys}.
	
	SLP is an approach that leverages both CSI and modulated data symbols to design the precoding strategy on a symbol-by-symbol basis. Unlike traditional block-level precoding (BLP), which treats interference as detrimental to system performance, the superiority of the SLP method lies in its ability to identify instantaneous interference into either constructive interference (CI)\cite{Selective-Precoding} or destructive interference (DI)\cite{DI}, and DI can be further converted into CI by the precoding design. As a further development, optimization-based SLP has been implemented to achieve enhanced communication performance. In \cite{CF-PSK} and \cite{CF-QAM}, the authors proposed closed-form SLP solutions for MU multiple-input single-output (MISO) downlink communication systems using phase shift keying (PSK) and quadrature amplitude modulation (QAM) constellation symbols, respectively, where the the bit error rate (BER) performance was improved. However, the precoding matrix must be optimized on a symbol-by-symbol basis, which poses a significant computational burden on the MU-MISO communication systems. To alleviate the computational costs, a low-complexity grouped SLP approach was proposed in \cite{Group SLP}, where a performance-complexity trade-off was achieved by transforming the intra-group interference into CI while suppressing the inter-group interference. Additionally, SLP method can be combined with hybrid precoding or low-resolution converter techniques for large-scale mmWave MIMO systems, leading to significant power savings \cite{Large-scale,DACs}. In contrast to the aforementioned SLP approaches that aim to maximize the CI effect, a more intuitive problem formulation is to minimize the symbol error rate (SER) \cite{TCOM SER,WCL SER}, wherein the SER expression has been derived and efficient algorithms have been proposed to deal with the non-convex problems.
	
	Notably, the majority of existing SLP researches primarily focus on MU-MISO downlink systems, i.e., with single antenna receivers, while only limited works addressing the challenges of MU-MIMO SLP problems, where the users employ multiple antennas \cite{CI Joint Combiner and Precoder,PSK SLP MU-MIMO,Con JTRSLP,Jour JTRSLP}. The key distinction in MIMO systems lies in the need for additional receive decoding techniques at the user side, particularly with the SLP method, where interference plays a different role compared to traditional approaches as it has been transformed into desired signals. The joint design of CI combiner and precoder was first considered in \cite{CI Joint Combiner and Precoder}, but only one data stream was transmitted for each user. In \cite{PSK SLP MU-MIMO}, the closed-form transmit SLP matrix and receive combining matrix were derived for MU-MIMO systems employing PSK modulated symbols, where a novel regularized interference rejection combiner (RIRC) receiver was proposed for signal decoding.  In \cite{Con JTRSLP,Jour JTRSLP}, the authors considered a joint SLP and linear receive combining design problem to minimize the SER for MU-MIMO systems using QAM constellation symbols. However, the dependency of the receive combining on the transmit signals was not fully addressed. Given the above analysis, SLP for MU-MIMO systems remains insufficiently explored, and therefore there is a need for more practical symbol-level schemes to be studied for MU-MIMO systems when using multi-level modulations such as QAM. 
	
	\subsection{Previous Work on Receive Detection}\label{PWORD}
	At the receiver side, MIMO detection techniques are employed to detect the received signals\cite{Fifty years mimo detection,LOW complxity MIMO Detection}. These techniques can be broadly classified into two types: linear and non-linear. Typical linear detection schemes offer lower computational complexity, such as ZF\cite{ZF Detection}, minimum mean-squared error (MMSE)\cite{MMSE Detection} and maximum asymptotic-multiuser-efficiency (MAME)\cite{MAME Detection}. However, the performance of these linear detection schemes deteriorates significantly compared to the optimal maximum likelihood detection (MLD) method\cite{MLD}. 
	
	MLD is a non-linear detection method that returns the optimal detection performance. However, the disadvantage of this method is that its computational complexity increases exponentially as the modulation order or the number of data streams increases. Given the increasing demand for achieving excellent transmission performance with low computational complexity, schemes that offer better performance-complexity tradeoff have been investigated \cite{QRMLD,QRMMLD}. A novel QR decomposition based MLD (QR-MLD) method was proposed in \cite{QRMLD}, where the channel matrix was transformed into an upper triangular matrix via QR decomposition, leading to a reduction in the number of signal candidates that need to be searched. Furthermore, the authors proposed the QR decomposition with \textit{M-algorithm} (QRM-MLD) method in \cite{QRMMLD}, which allowed for a further complexity reduction. Although the QRM-MLD method significantly decreases the decoding computational complexity compared to the MLD and QR-MLD methods, this improvement comes at the expense of  decoding accuracy.
	
	Based on the aforementioned analysis, the MLD method is applied at the user side to decode the received signals in this paper for the following reasons: 1) MLD method provides promising performance, and an efficient performance-complexity trade-off can be achieved with the alternative QRM-MLD method; 2) This method enhances the generality of our proposed SLP schemes, as it requires no modification to receiver architectures; 3) Both MLD and QRM-MLD methods have been widely adopted in commercial devices, thereby increasing the practicality of the SLP schemes proposed in this paper \cite{Emld}.
	
	\subsection{Contributions}\label{Contri}
	In this paper, we design practical symbol-level transmit precoding and receive decoding techniques for MU-MIMO communication systems. We summarize the main contributions of this paper as follows:
	\begin{enumerate}
		\item We first formulate a joint symbol-level transmit precoding and receive combining optimization problem for MU-MIMO systems with QAM modulated data symbols, where we maximize the CI effect of the outer constellation symbols while maintaining the performance of the inner constellation symbols. To solve the coupled optimization problem, we utilize the alternating optimization (AO) method to decouple the variables, transforming the original problem into two convex subproblems.
		\item By analyzing the Lagrangian and KKT conditions, and applying mathematical analysis to the subproblems, we derive the optimal transmit precoding and receive combining matrix structures in closed forms as functions of the dual variables. These subproblmes are further simplified into QP problems that can be solved in low complexity. Additionally, the convergence and computational complexity of the proposed AO algorithm are analyzed.
		\item To address the dependence of the receive combining matrix on the transmit symbols, we further utilize the MLD method at the user side to decode the received signals. Importantly, we demonstrate that the performance of MLD method deteriorates as the smallest singular value of the transmit precoding matrix decreases. Furthermore, we have shown that the traditional SLP matrix is rank-one, making it infeasible to directly apply the MLD method at the users. Additionally, we highlight that the SLP method presents challenges for independent MLD decoding for each user. Thus, it is essential to incorporate independent decoding constraint into the optimization problem.
		\item We propose a novel optimization problem to maximize the lower bound of the smallest singular value of the transmit SLP matrix, subject to the CI constraints, independent decoding requirement and the transmit power budget. Moreover, we derive a generic semi-definite programming (SDP)-based problem to reduce the number of variables to be optimized by omitting the CI constraints and focusing solely on optimizing the smallest singular value of the transmit precoding matrix.
	\end{enumerate}
	
	Simulation results demonstrate that the proposed schemes achieve significant performance improvements over conventional BD-based approach. Our proposed singular value optimization problems outperform the joint design scheme, which benefits from the superiority of MLD over linear detection algorithms. The SDP-based problem experiences only a minor performance loss, but benefits from fewer variables need to be optimized.
	
	\subsection{Organization and Notations}\label{OAN}
	The rest of this paper is organized as follows. Section \ref{SMPF} introduces the MU-MIMO system model under consideration, provides a brief review of the CI metric for QAM constellation, and presents the corresponding formulated problem. Section \ref{JSLTARBD} provides an analysis of the joint symbol-level transmit precoding and receive combining optimization problem, and derives the structures of the closed-form solutions. Section \ref{SLPWMLD} analyzes the performance of the MLD method and the properties of the SLP matrix, and proposes two symbol-level singular value optimization problems to enhance decoding efficiency. In Section \ref{CCCA}, the convergence and computational complexity are analyzed for proposed algorithms. The simulation results are shown in Section \ref{Simulation}, and Section \ref{Conclusion} concludes the paper.
	
	Notations: a, $ \bf a $ and $ \bf A $ denote scalar, vector, and matrix, respectively.  $\mathbb{C}^{M\times N}$ or $\mathbb{R}^{M\times N}$ ($\mathbb{C}^{M \times 1}$ or $\mathbb{R}^{M \times 1}$) represents a complex-valued or real-valued  $ M \times N $ matrix ($ M \times 1 $ vector). The transpose, conjugate and complex conjugate transpose of a matrix or vector are denoted by using $(\cdot)^T$, $(\cdot)^*$ and $(\cdot)^H$. The inverse and trace of a matrix are $(\cdot)^{-1}$ and $\text{Tr}(\cdot)$. $a(i,j) $ is the entry in the $i$-row and $j$-th column of matrix $\bf{A}$ and $\Vert \cdot \Vert_2$ represents the ${l_2}$-norm. $ \frak{R}(\cdot) $ and $ \frak{J}(\cdot) $ denote the real and imaginary part of a complex number, respectively. $card\{\cdot\}$ denotes the cardinality of a set, and $\otimes$ is the Kronecker product. $j$ represents the imaginary unit, ${\bf{I}}_K$ is the $K \times K$ identity matrix, and ${\bf e}_i$ denotes the \textit{i}-th column of the identity matrix. $\text{diag}(\bf a)$ denotes a square diagonal matrix with the elements of vector $\bf a$ on the main diagonal.

	\section{System Model and Problem Formulation}\label{SMPF}
	
	\subsection{System Model}\label{SysMod}
	We study a downlink MU-MIMO system, where the BS is equipped with ${N_T}$ transmit antennas and simultaneously transmits $L$ data streams to each user. Each user is equipped with ${N_R}$ antennas, with the constraint $L \le N_R$. The total number of transmit antennas satisfies ${N_T} \ge KL$, where $K$ is the number of users served by the BS. We focus on the MU-MIMO transceiver design, where perfect CSI is assumed throughout the paper\cite{Con JTRSLP,Jour JTRSLP,PSK SLP MU-MIMO}. The transmit symbol vector is assumed to be formed from a normalized QAM constellation\cite{CF-QAM}, and ${{\bf{s}}_k} \in \mathbb{C} {^{L \times 1}}$ represents the transmit signal for the $k$-th user. The transmit precoding matrix for the $k$-th user is denoted as ${{\bf{P}}_k} \in \mathbb{C} {^{N_T \times L}}$. Therefore, the transmit signal at the BS can be expressed as
	\begin{equation}\label{transmit signal}
		{\bf{x}} = {\bf{Ps}} = \sum\limits_{k = 1}^K {{{\bf{P}}_k}{{\bf{s}}_k}}, 
	\end{equation}
	where ${\bf{P}} =\left[ {{{\bf{P}}_1},{{\bf{P}}_2}, \cdots ,{{\bf{P}}_K}} \right]\in \mathbb{C} {^{N_T \times KL}}$ and ${\bf{s}} = {\left[ {{\bf{s}}_1^T,{\bf{s}}_2^T, \cdots ,{\bf{s}}_K^T} \right]^T}\in \mathbb{C} {^{KL\times 1}}$ represent the overall transmit precoding matrix and symbol vector, respectively. Then, the received signal at the $k$-th user is given by
	\begin{equation}\label{receive signal}
		\begin{array}{l}
			{{\bf{y}}_k} = {{\bf{H}}_k}{\bf{x}} + {{\bf{n}}_k}  \vspace{1ex}\\
			{\kern 11pt}  = {{\bf{H}}_k}{{\bf{P}}_k}{{\bf{s}}_k} + {{\bf{H}}_k}\sum\limits_{i = 1,i \ne k}^K {{{\bf{P}}_i}{{\bf{s}}_i}}  + {{\bf{n}}_k}, \\
%			{\kern 11pt} = {{\bf{G}}_k}{{\bf{s}}_k} + \sum\limits_{i = 1,i \ne k}^K {{{\bf{G}}_{k,i}}{{\bf{s}}_i}}  + {{\bf{n}}_k},
		\end{array}
	\end{equation}
	where ${\bf{H}}_k \in \mathbb{C}{^{{N_R} \times {N_T}}}$ denotes the channel matrix between the \textit{k}-th user and the BS, ${{\bf{n}}_k} \in \mathbb{C}{^{N_R \times 1}}  $ represents the zero mean circularly symmetric complex Gaussian noise vector with ${{\bf{n}}_k} \sim \mathcal{CN}\left( {{\bf{0}},{\sigma ^2}{\bf{I}}} \right)$. To correctly decode the received signals, the $k$-th user decodes the received signal with a receive combining matrix ${{\bf{W}}_k} \in \mathbb{C} {^{L \times N_R}}$, then the signal for demodulation can be expressed as
	\begin{equation}\label{decoded signal} 
		\begin{array}{l}
			{{{\bf{\hat s}}}_k} = {{\bf{W}}_k}\left( {{{\bf{H}}_k}\sum\limits_{i = 1}^K {{{\bf{P}}_i}{{\bf{s}}_i}}  + {{\bf{n}}_k}} \right) \vspace{1ex}\\
			{\kern 9pt}  = {{\bf{W}}_k}{{\bf{H}}_k}{{\bf{P}}_k}{{\bf{s}}_k} + {{\bf{W}}_k}{{\bf{H}}_k}\sum\limits_{i = 1,i \ne k}^K {{{\bf{P}}_i}{{\bf{s}}_i}}  + {{\bf{W}}_k}{{\bf{n}}_k},
		\end{array}
	\end{equation}
	
	\subsection{Constructive Interference}\label{CI}
	\begin{figure}[t]
		\centering 
		\subfigure[]{
			\label{figureb}
			\includegraphics[scale=0.25]{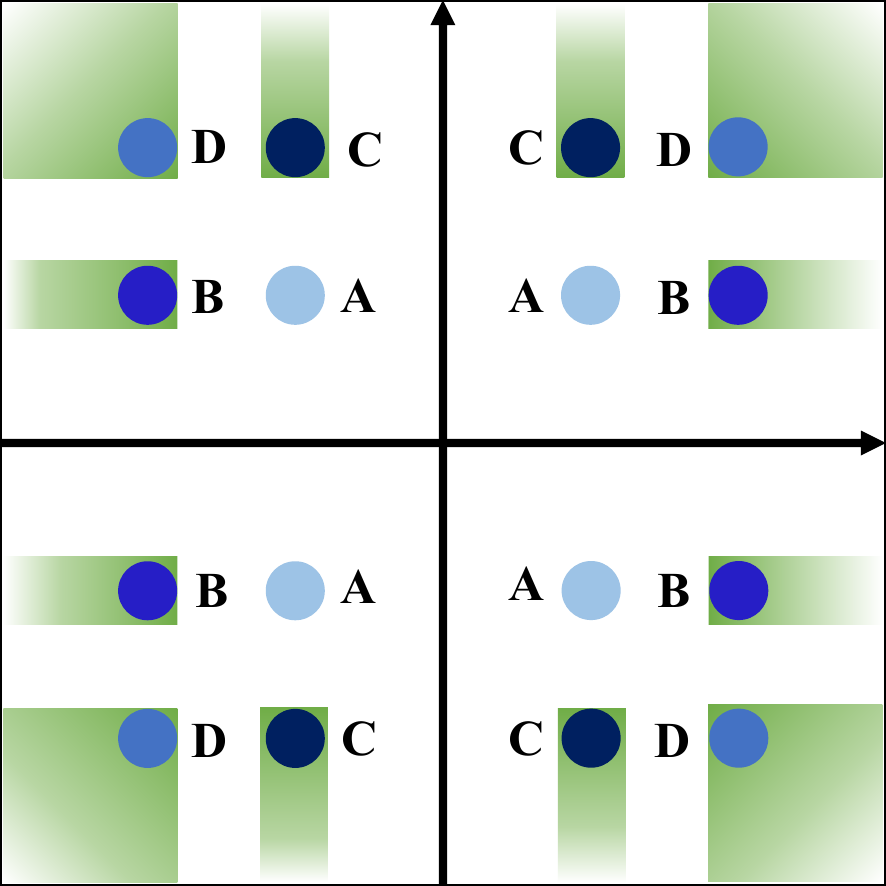}}
		\subfigure[]{
			\label{figurea}
			\includegraphics[scale=0.52]{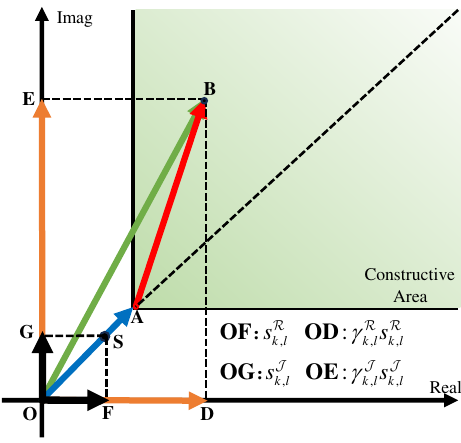}}
		\caption{Constellation point categorization and CI metric for 16QAM.}
		\label{moxing}
	\end{figure}
	To enhance the understanding of the subsequent problem formulation, a concise review of CI metric for QAM modulations is provided. For illustration, Fig. \ref{figureb} shows the entire constellation map of the 16QAM constellation, and Fig. \ref{figurea} depicts one quadrant of a nominal 16QAM constellation in detail. It shows that the constellation points can be divided into 4 types \cite{interference1}
	\begin{enumerate}
		\item Type `A': inner points that no CI can be exploited;
		\item Type `B': outer points that CI can be exploited for the real part;
		\item Type `C': outer points that CI can be exploited for the imaginary part;
		\item Type `D': outer points that CI can be exploited for both the real and imaginary parts.
	\end{enumerate}
	Accordingly, the `symbol-scaling' metric discussed in \cite{CF-QAM} is presented here. Specifically, the constellation points and noiseless received signals can be mathematically decomposed into
	\begin{equation*}\label{signal decomposition}
	\begin{aligned}
			{s_{k,l}} & = s_{k,l}^{\cal R} + s_{k,l}^{\cal J} \vspace{1ex} \\
			{{\bf{w}}_{k,l}}{{\bf{H}}_k}{\bf{Ps}} &= \gamma _{k,l}^{\cal R}s_{k,l}^{\cal R} + \gamma _{k,l}^{\cal J}s_{k,l}^{\cal J},\forall k \in {\cal K},\forall l \in {\cal L},
	\end{aligned}
	\end{equation*}
	where ${s_{k,l}}$ denotes the \textit{l}-th data symbol for the \textit{k}-th user, $s_{k,l}^{\cal R} = \mathfrak{R} \left\{ {{s_{k,l}}} \right\}$ and $s_{k,l}^{\cal J} = j \cdot {\frak J}\left\{ {{s_{k,l}}} \right\}$ are the bases that are parallel to the detection thresholds for each constellation symbol. ${{\bf{w}}_{k,l}}$ is the \textit{l}-th row of the receive combining matrix ${{\bf{W}}_k}$. $\gamma _{k,l}^{\cal R} \ge 0$ and $\gamma _{k,l}^{\cal J} \ge 0$ are the real-valued scaling coefficients, and a larger value of $\text{min}\left\{ \gamma _{k,l}^{\cal R},\gamma _{k,l}^{\cal J} \right\}$ indicates a larger distance to the decision boundaries, leading to a better BER performance. To concisely express the CI constraints, we define
	\begin{equation}\label{gamma&s}
	\begin{array}{l}
		{{\bm{\gamma }}_{k,l}} = {\left[ {\gamma _{k,l}^{\cal R},\gamma _{k,l}^{\cal J}} \right]^T}$, ${{{\bf{\bar s}}}_{k,l}} = {\left[ {s_{k,l}^{\cal R},s_{k,l}^{\cal J}} \right]^T},\forall k \in {\cal K},\forall l \in {\cal L},
	\end{array}
	\end{equation}
	then ${{\bf{w}}_{k,l}}{{\bf{H}}_k}{\bf{Ps}}$ can be further simplified as
	\begin{equation}
		{{\bf{w}}_{k,l}}{{\bf{H}}_k}{\bf{Ps}} = {\bm{\gamma }}_{k,l}^T{{{\bf{\bar s}}}_{k,l}},\forall k \in {\cal K},\forall l \in {\cal L}.
	\end{equation}
	
	For notational convenience, we introduce a set $\mathcal{O}$ that consists of the real or imaginary parts of the outer points that can be scaled, and a set $\mathcal{I}$ that consists of the real or imaginary parts of the inner points that can not be scaled\cite{CF-QAM}. Accordingly, the set $\mathcal{O} \cup \mathcal{I}$ includes all constellation points, and the CI constraints for QAM symbols can be expressed as
	\begin{equation}
	\gamma _{k,l}^\mathcal{O} \ge t,\gamma _{k,l}^\mathcal{I} = t,\forall \gamma _{k,l}^\mathcal{O} \in \mathcal{O},\forall \gamma _{k,l}^\mathcal{I} \in \mathcal{I},
	\end{equation}
	where $t$ is the Euclidean distance between the CI region and decision boundary, which is the objective to be maximized. A more detailed description can be found in \cite{SLP Surveys,CF-QAM}.
	 
	\subsection{Problem Formulation}\label{ProFor}
	
	Consistent with \cite{CF-QAM}, we consider interference on the inner points as destructive and on the outer points as constructive. Following \cite{PSK SLP MU-MIMO}, we propose to jointly optimize the transmit precoding matrix and receive combining matrix to maximize the CI effect for the outer constellation points while maintaining performance for the inner constellation points. For the considered MU-MIMO system, the problem can be formulated as
	\begin{equation}\label{P1}
		\begin{array}{l}
	 		{{\text{P1}}}:{\kern 2pt}\mathop {\max }\limits_{{{\bf{W}}_k},{\bf{P}},t,{{\bf{\Gamma }}}} {\kern 4pt} t \vspace{1ex} \\
	 		{\kern 35pt}s.t.{\kern 18pt}{\bf{C1}}:{{\bf{W}}_k}{{\bf{H}}_k}{\bf{Ps}} = {{\bf{\Gamma }}_k}{{{\bf{\bar s}}}_k},{\kern 1pt} \forall k \in {\cal K} \vspace{1ex} \\
	 		{\kern 67pt} {\bf{C2}}:t - \gamma _{m}^{\cal O} \le 0,\forall \gamma _{m}^{\cal O} \in {\cal O} \vspace{1ex} \\
	 		{\kern 67pt} {\bf{C3}}:t - \gamma _{n}^{\cal I} = 0,\forall \gamma _{n}^{\cal I} \in {\cal I} \vspace{1ex} \\
	 		{\kern 67pt} {\bf{C4}}:\left\| {{\bf{Ps}}} \right\|_2^2 \le p \vspace{1ex} \\
	 		{\kern 67pt} {\bf{C5}}:\left\| {{{\bf{W}}_k}} \right\|_2^2 \le 1,\forall k \in {\cal K},
	 	\end{array}
	\end{equation}
	where ${{\bm{\Gamma }}_k} = {\left[ {{\bm{\gamma }}_{k,1}^T,{\bm{\gamma }}_{k,2}^T, \cdots ,{\bm{\gamma }}_{k,L}^T} \right]^T}$ is the real-valued scaling vector and ${{\bf{\bar s}}_k} = {\left[ {{\bf{\bar s}}_{k,1}^T,{\bf{\bar s}}_{k,2}^T, \cdots ,{\bf{\bar s}}_{k,L}^T} \right]^T}$ is the transmit signal basis vector for the \textit{k}-th user. The objective function of the formulated problem is the distance between the CI region and decision boundary. {\bf{C1}}-{\bf{C3}} represent the CI constraints for QAM symbols, and we can obtain
	\begin{equation}
		\begin{array}{l}
		{\cal O} \cup {\cal I} = \left\{ {\gamma _{1,1}^{\cal R},\gamma _{1,1}^{\cal J}, \cdots ,\gamma _{k,l}^{\cal R},\gamma _{k,l}^{\cal J}, \cdots ,\gamma _{K,L}^{\cal R},\gamma _{K,L}^{\cal J}} \right\},  \vspace{1ex} \\
		card\left\{ {\cal O} \right\} + card\left\{ {\cal I} \right\} = 2KL.
		\end{array}
	\end{equation}
	{\bf{C4}} is the total transmit power budget and {\bf{C5}} is the receive combining power budget to ensure the problem is bounded and feasible. We should note that the value of the power constraint at the right-hand side of {\bf{C5}} will not affect the communication performance, because ${{\bf{W}}_k}$ is multiplied to the noise, too.
	
	\section{Joint Symbol-Level Transmit Precoding and Receive Combining Design} \label{JSLTARBD}
	
	In this section, we focus on solving the joint symbol-level transmit precoding and receive combining design problem formulated in (\ref{P1}). The AO algorithm is employed to iteratively optimize the transmit precoding matrix and the receive combining matrix. Specifically, ${{\text{P1}}}$ can be transformed into two subproblems by fixing ${\bf{P}}$ or ${{\bf{W}}_k}$, allowing for more efficient algorithms to separately optimize the subproblems with simpler structures.
	
	\subsection{Optimization on Transmit Precoding Matrix}\label{OptimizeP}
	
	We first optimize the transmit precoding matrix ${\bf{P}}$ for a given receive combining matrix ${{\bf{W}}_k}$. Fortunately, the formulated subproblem takes a similar form to the optimized problem in \cite{CF-QAM}, and the closed-form solution can be obtained by analyzing the Lagrangian function and KKT conditions. Furthermore, the modified iterative algorithm proposed in \cite{CF-QAM} can also be applied to obtain the optimal precoding matrix and accelerate the convergence speed of the proposed AO algorithm. To be more specific, when ${{\bf{W}}_k}$ is fixed, ${{\text{P1}}}$ can be transformed into 
	\begin{equation}\label{P2}
		\begin{array}{l}
			{{\text{P2}}}:{\kern 1pt}\mathop {\max }\limits_{{\bf{P}},t,{\bf{\Gamma }}} {\kern 4pt} t \vspace{1ex}\\
			{\kern 25pt}  s.t. {\kern 5pt} {\bf{C1}}:{\bf{GPs}} = {\bf{U}}\text{diag}\left( {\bf{\Gamma }} \right){{\bf{{\bar s}}}} \vspace{1ex}\\
		    {\kern 43pt} {\bf{C2}}:t - \gamma _{m}^{\cal O} \le 0,\forall \gamma _{m}^{\cal O} \in {\cal O} \vspace{1ex}\\
			{\kern 43pt}{\bf{C3}}:t - \gamma _{n}^{\cal I} = 0,\forall \gamma _{n}^{\cal I} \in {\cal I} \vspace{1ex}\\
			{\kern 43pt} {\bf{C4}}:\left\| {{\bf{Ps}}} \right\|_2^2 \le p,
		\end{array}
	\end{equation}
	where ${\bf{G}} = {\bf{WH}}$, ${\bf{H}} = {\left[ {{\bf{H}}_1^T,{\bf{H}}_2^T,\cdots,{\bf{H}}_K^T} \right]^T}$ is the channel matrix and $\bf{W}$ is a diagonal matrix composed of ${{\bf{W}}_k}$ and expressed as
	\begin{equation}\label{W}
	{\bf{W}} = \left[ {\begin{array}{*{20}{c}}
				{{{\bf{W}}_1}}&{\bf{0}}& \cdots &{\bf{0}}\\
				{\bf{0}}&{{{\bf{W}}_2}}&{\bf{0}}& \vdots \\
				\vdots &{\bf{0}}& \ddots &{\bf{0}}\\
				{\bf{0}}& \cdots &{\bf{0}}&{{{\bf{W}}_K}}
		\end{array}} \right].
	\nonumber 
	\end{equation}
	${{\bf{\bar s}}} = {\left[ {{\bf{\bar s}}_1^T,{\bf{\bar s}}_2^T,\cdots,{\bf{\bar s}}_K^T} \right]^T}$, ${\bf{\Gamma }} = {\left[ {{\bf{\Gamma }}_1^T,{\bf{\Gamma }}_2^T, \cdots ,{\bf{\Gamma }}_k^T} \right]^T}$ and ${\bf{U}} = {{\bf{I}}_{KL}} \otimes \left[ {1,1} \right]$. The closed-form solution of P2 can be expressed as\cite{CF-QAM}
	\begin{equation}\label{P}
		\begin{array}{l}
	{\bf{P}} \vspace{1ex}\\= \dfrac{1}{{KL}}{{\bf{G}}^H}{\left( {{\bf{G}}{{\bf{G}}^H}} \right)^{ - 1}}{\bf{U}}\text{diag}\left( {\sqrt {\dfrac{p}{{{{\bf{u}}^T}{{{\bf{\tilde V}}}^{ - 1}}{\bf{u}}}}} {{\bf{E}}^{ - 1}}{{{\bf{\tilde V}}}^{ - 1}}{\bf{u}}} \right){{\bf{{\bar s}}}}{{\bf{\hat s}}^T},
		\end{array}
	\end{equation}
	where ${\bf{\hat s}} = \left[ {\dfrac{1}{{{s_1}}},\dfrac{1}{{{s_2}}}, \cdots ,\dfrac{1}{{{s_{KL}}}}} \right]$, ${\bf{u}}$ is the dual variable vector and expressed as
	\begin{equation}
	{\bf{u}} = {\left[ {{\mu _1},{\mu _2}, \cdots ,{\mu _{card\left\{ O \right\}}},{\nu _1},{\nu _2}, \cdots ,{\nu _{card\left\{ I \right\}}}} \right]^T},
	\end{equation}
	which can be obtained by solving a simpler QP optimization problem P3, given by
	\begin{equation}\label{P3}
		\begin{array}{l}
			{{\text{P3}}}:\mathop {{\rm{min}}}\limits_{\bf{u}}{\kern 4pt} {{\bf{u}}^T}{{{\bf{\tilde V}}}^{ - 1}}{\bf{u}}\vspace{1ex}\\
			{\kern 21pt} s.t.{\kern 7pt} {\bf{C1}}:{{\bf{1}}^T}{\bf{u}} - 1 = 0\vspace{1ex}\\
			{\kern 42pt} {\bf{C2}}:{\mu _m} \ge 0,\forall m \in \left\{ {1,2,\cdots,card\left\{ \cal{O} \right\}} \right\},
		\end{array}
	\end{equation}
	where ${\bf{\tilde V}} = {\bf{EV}}{{\bf{E}}^T}$, ${\bf{V}} = \frak{R} \left\{ {\bf{T}} \right\}$, and ${\bf{T}}$ and ${\bf{E}}$ are expressed as
	\begin{subequations}
	\begin{align}
		&{\bf{T}} = \text{diag}\left( {{\bf{\bar s}}^H} \right){{\bf{U}}^H}{\left( {{\bf{G}}{{\bf{G}}^H}} \right)^{ - 1}}{\bf{U}}\text{diag}\left( {{{\bf{{\bar s}}}}} \right), \vspace{1ex}\\  
		&{\bf{E}} = { \left[ {\bf{e}}_{L ({{\tilde s}_1})}, {\bf{e}}_{L ({{\tilde s}_2})} , \cdots , {\bf{e}}_{L ({{\tilde s}_{2KL}})} \right]^T}.
	\end{align}
	\end{subequations}
	Here, $ L(\cdot) $ is a `\textbf{Locater}' function that returns the index of ${{\tilde s}_m}$ in ${\bf{{\bar s}}}$, defined as
	\begin{equation}
		L({\tilde s}_m) = k, {\kern 2pt}\text{if}{\kern 4pt} {\tilde s}_m = {\bar s}_k.
	\end{equation}
	${\bf{E}}\in \mathbb{R} {^{2KL \times 2KL}}$ is an invertible matrix introduced to rearrange the columns and rows of the matrix ${\bf{ V}}$ for notational and mathematical convenience\cite{CF-QAM}.
	
	\subsection{Optimization on Receive Combining Matrix}\label{OptimizeW}
	After obtaining the transmit precoding matrix ${\bf{P}}$, the next step is to optimize the receive combining matrix ${{\bf{W}}_k}$, where a total number of $K$ combining matrices can be optimized in parallel. Specifically, given a known ${\bf{P}}$,  the optimization on a specific ${{\bf{W}}_k}$ can be reformulated as
	\begin{equation}
	\begin{array}{l}
			{{\text{P4}}}:\mathop {\max }\limits_{{{\bf{W}}_k},t,{\bf{\Gamma }}}{\kern 4pt} t \vspace{1ex}\\
			{\kern 27pt} s.t.{\kern 7pt} {\bf{C1}}:{{\bf{W}}_k}{{\bf{r}}_k} = {{\bf{U}}_1}\text{diag}\left( {{{\bf{\Gamma }}_k}} \right){{{\bf{\bar s}}}_k},\forall k \in {\cal K} \vspace{1ex}\\
			{\kern 48pt} {\bf{C2}}:t - \gamma _{m}^{\cal O} \le 0,\forall \gamma _{m}^{\cal O} \in {\cal O} \vspace{1ex}\\
			{\kern 48pt} {\bf{C3}}:t - \gamma _{n}^{\cal I} = 0,\forall \gamma _{n}^{\cal I} \in {\cal I} \vspace{1ex}\\
			{\kern 48pt}{\bf{C4}}:\left\| {{{\bf{W}}_k}} \right\|_2^2 \le 1,\forall k \in {\cal K},
	\end{array}
	\end{equation}
	where ${{\bf{r}}_k} = {{\bf{H}}_k}{\bf{Ps}}$, ${\bf{U}}_1 = {{\bf{I}}_{L}} \otimes \left[ {1,1} \right]$. P4 is a second-order-cone programming (SOCP) problem and solved readily. Furthermore, the optimal ${{\bf{W}}_k}$ is shown to exhibit a closed-form expression based on the following proposition.
	
	\textit{Proposition 1}: The optimal receive combining matrix ${{\bf{W}}_k}$ for P4 can be expressed as
	\begin{equation}\label{OptimalWk}
		{{\bf{W}}_k} = \frac{1}{{{\bf{r}}_k^H{{\bf{r}}_k}}}{{\bf{U}}_1}\text{diag}\left( {\sqrt {\frac{{{\bf{r}}_k^H{{\bf{r}}_k}}}{{{\bf{u}}_1^T{\bf{\tilde V}}_1^{ - 1}{{\bf{u}}_1}}}} {\bf{E}}^{ - 1}{\bf{\tilde V}}_1^{ - 1}{{\bf{u}}_1}} \right){{\bf{\bar s}}_k}{\bf{r}}_k^H,
	\end{equation}
	where ${{\bf{\tilde V}}_1} = {\bf{E}}{{\bf{V}}_1}{{\bf{E}}^T}$, ${{\bf{V}}_1} = {\frak R}\left( {{{\bf{T}}_1}} \right)$ and ${{\bf{T}}_1} = \text{diag}\left( {{\bf{\bar s}}_k^H} \right){\bf{U}}_1^H{{\bf{U}}_1}\text{diag}\left( {{{{\bf{\bar s}}}_k}} \right)$. 
	
	\textit{Proof}: See Appendix A.
	
	Furthermore, ${{\bf{u}}_1}$ is obtained by solving the following QP optimization problem:
	\begin{equation}\label{P6}
		\begin{array}{l}
			{\text{P5}}:{\kern 1pt}\mathop {\min }\limits_{{{\bf{u}}_1}} {\kern 5pt} {\bf{u}}_1^T{\bf{\tilde V}}_1^{ - 1}{{\bf{u}}_1}  \vspace{1ex}\\
			{\kern 23pt}  s.t.{\kern 7pt} {\bf{C1}}: {\kern 1pt} {{\bf{1}}^T}{{\bf{u}}_1} - 1 = 0  \vspace{1ex}\\
			{\kern 44pt} {\bf{C2}}:{\kern 1pt} {\mu _m} \ge 0,\forall m \in \left\{ {1,2, \cdots ,card\left\{ {\cal O} \right\}} \right\}.
		\end{array}
	\end{equation}
	The detailed derivation process of P5 is similar to that of ${\mathcal{P}}_5$ in \cite{CF-QAM}, which can be found in \textit{Proposition} 2 presented there. Furthermore, P5 is a convex QP problem in the same form as P3 and can be effectively solved.
	
	\subsection{Alternating Optimization Algorithm}\label{IRC1}
	\begin{algorithm}[!b]
		\caption{Proposed AO algorithm for solving problem (\ref{P1})}
		\label{al1}
		\begin{algorithmic}
			\State ${\bf Input:}$ ${\bf{H}}$, ${\bf{s}}$, and convergence threshold $\kappa$.\vspace{0.5ex}
			\State ${\bf Output:}$ ${\bf{P}}$, ${\bf{W}}_k$.\vspace{0.5ex}
			\State ${\bf Initialization:}$  Initialize ${\bf{P}}^{(1)}$, $t^{(1)}=0$, $ \kappa=10{^{-5}}$, and iteration index $ \textit{n}=1 $.\vspace{0.5ex}
			\While {$\delta > \kappa $} \vspace{0.5ex}
			\State Calculate ${\bf{Q}}_1 = {{{\bf{\tilde V}}_1}^{ - 1}}$ \vspace{0.5ex}
			\State Obtain ${\bf{ u}}_1$ via Algorithm 1 proposed in \cite{CF-QAM} \vspace{0.5ex}
			\State Obtain ${{\bf{W}}_{k}^{(n+1)}} $  via (\ref{Wk})\vspace{0.5ex}
			\State Calculate ${\bf{Q}} = {{{\bf{\tilde V}}}^{ - 1}}$ \vspace{0.5ex}
			\State Obtain ${{\bf{ u}}}$ via Algorithm 1 proposed in \cite{CF-QAM} \vspace{0.5ex}
			\State Obtain ${{\bf{P}}^{(n+1)}}$ and $t^{(n+1)}$ via (\ref{P}) and (\ref{P2}) \vspace{0.5ex}
			\State $\delta  = \left| t^{(n+1)}-t^{(n)} \right|$
			\State $n=n+1$
			\EndWhile
			\State ${\bf P}={\bf P}^{\left( {n} \right)}$, ${{\bf{W}}_k}={\bf{W}}_k^{\left( {n} \right)}$.
		\end{algorithmic}
	\end{algorithm}
	
	The optimal solutions to P1 can be obtained by iteratively solving the QP problems P3 and P5, until convergence. It can be observed that the number of variables to be optimized in P2, P3, P4 and P5 are ${N_T}KL$, $2KL$, ${N_R}KL$ and $2KL$, respectively, which demonstrates that the computational complexity of the two subproblems is reduced. Furthermore, P3 and P5 exhibit the same form as ${\mathcal{P}}_5$ in \cite{CF-QAM}, which indicates the generic iterative algorithm proposed therein also can be used to update our variables, further accelerating the solving process. Based on the preceding discussion, we present the proposed AO algorithm for the joint precoding and combining problem P1, as summarized in \textbf{Algorithm \ref{al1}}.

	In our proposed AO algorithm, the transmit precoding matrix ${\bf{P}}$ and receive combining matrix ${\bf{W}}_k$ are iteratively updated until convergence. Furthermore, the iterative algorithm introduced in \cite{CF-QAM} is used to effectively solve the two QP subproblems to obtain the dual variables. An additional analysis of convergence and computational complexity for the proposed algorithm is presented in Section V.

	\section{Symbol-Level Precoding for MLD Decoding} \label{SLPWMLD}
	Although the aforementioned joint symbol-level transmit precoding and receive combining optimization scheme can achieve excellent communication performance, the reliance of the receive combining matrix on the transmit data symbols makes it inapplicable in practice. To address this challenge and meanwhile ensure promising decoding performance, in this section we consider the SLP design at the BS when users employ the MLD decoding method. We first show that the traditional SLP design is not applicable in such case because the rank-one property of the SLP matrix, followed by the proposition of a novel SLP solution tailored for MLD that offers a promising performance.
	
	\subsection{Performance Analysis of MLD Method}\label{PAoMLD}
	The MLD method is a non-linear detection scheme and is regarded as the optimal signal detection method for MIMO systems\cite{LOW complxity MIMO Detection}. However, if the transmit precoding matrix is rank-deficient, the performance of MLD will deteriorate severely. In this section, we demonstrate that the smallest singular value of the precoding matrix is a crucial factor that affects the system performance when users employ MLD as the decoding method, which motivates our subsequent design.
	
	In a generic MU-MIMO system where users employ MLD method to decode the received signals, the estimation process can be expressed as
	\begin{equation}\label{estimation}
	\begin{array}{l}
		{\bf{\hat s}} = \mathop {\arg }\limits_{\bf{\tilde s}} {\kern 1pt} \min \left\| {{\bf{y}} - {\bf{M\tilde s}}} \right\|_2^2 \vspace{1ex}\\
		{\kern 5pt}   = \mathop {\arg }\limits_{\bf{\tilde s}} {\kern 1pt}  \min \left\| {{\bf{M}}\left( {{\bf{s}} - {\bf{\tilde s}}} \right) + {\bf{n}}} \right\|_2^2 \vspace{1ex}\\
		{\kern 5pt} \mathop  = \limits^{(a)}  \mathop {\arg }\limits_{\bf{\tilde s}} {\kern 1pt}\min \left\{ {{\rm{Tr}}\left[ {\left( {{\bf{s}} - {\bf{\tilde s}}} \right){{\left( {{\bf {s}} - {\bf{\tilde s}}} \right)}^H}{{\bf{M}}^H}{\bf{M}}} \right] + {\sigma ^2}} \right\} \vspace{1ex}\\
		{\kern 5pt}   = \mathop {\arg }\limits_{\bf{\tilde s}} {\kern 1pt}  \min \left\{ {{\rm{Tr}}\left[ {{\bf{\tilde S\tilde M}}} \right] + {\sigma ^2}} \right\},
	\end{array}
	\end{equation}
	where the subscript \textit{k} for the \textit{k}-th user is omitted for clarity. Here, ${\bf{M}} = {{\bf{H}}}{\bf{P}}$ denotes the equivalent transmit-receive channel matrix, ${\bf{\tilde s}}$ represents the candidate symbol vector selected from the constellation book. Additionally, we define ${\bf{\tilde S}} = \left( {{\bf{s}} - {\bf{\tilde s}}} \right){\left( {{\bf{s}} - {\bf{\tilde s}}} \right)^H}$ and ${\bf{\tilde M}} = {{\bf{M}}^H}{\bf{M}}$. The step $(a)$ can be achieved because the transmit symbols and received noise are independent. Furthermore, we introduce \textit{Von Neumann's trace inequality} here for subsequent derivation.
	
	\textit{Lemma 1} (\textit{Von Neumann's trace inequality}): Let $\bf{A}$ and $\bf{B}$ be the $N$-dimensional Hermitian positive semi-definite matrices. Denote the eigenvalues of $\bf{A}$ and $\bf{B}$ as ${\lambda _1}\left( {\bf{A}} \right) \ge {\lambda _2}\left( {\bf{A}} \right) \ge  \cdots  \ge {\lambda _N}\left( {\bf{A}} \right)$ and ${\lambda _1}\left( {\bf{B}} \right) \ge {\lambda _2}\left( {\bf{B}} \right) \ge  \cdots  \ge {\lambda _N}\left( {\bf{B}} \right)$, respectively, which are arranged in a non-decreasing order. Then we have
	\begin{equation}
	{\rm{Tr}}\left[ {{\bf{AB}}} \right] \ge \sum\limits_{n = 1}^N {{\lambda _n}\left( {\bf{A}} \right){\lambda _{N - n + 1}}\left( {\bf{B}} \right)}.
	\end{equation}
	
	Based on \textit{Lemma 1}, we obtain:
	\begin{equation}
	\begin{array}{l}
	{\rm{Tr}}\left[ {{\bf{\tilde S\tilde M}}} \right] \ge \sum\limits_{l = 1}^L {{\lambda _l}\left( {{\bf{\tilde S}}} \right){\lambda _{L - l + 1}}\left( {{\bf{\tilde M}}} \right)} \mathop  = \limits^{\left( b \right)} {\lambda _1}\left( {{\bf{\tilde S}}} \right){\lambda _L}\left( {{\bf{\tilde M}}} \right),
	\end{array}
	\end{equation}
	where the step $(b)$ follows from the fact that ${{rank}}({\bf{\tilde S}}) = 1$. According to (\ref{estimation}), under the same SNR conditions, a larger value of ${\rm{Tr}}( {{\bf{\tilde S\tilde M}}} )$ is more beneficial for correctly decoding the transmit symbols from the received signals that contain noise, which can be achieved by maximizing ${\lambda _1}({{\bf{\tilde S}}}) {\lambda _L}({{\bf{\tilde M}}})  $, the lower bound of ${\rm{Tr}}( {{\bf{\tilde S\tilde M}}} )$. Moreover, ${\lambda _1}( {{\bf{\tilde S}}} )$ remains constant once the candidate symbol vector is selected, and since ${\bf{\tilde M}}$ is comprised of the optimized transmit precoding matrix ${\bf{P}}$, the singular values of ${\bf{P}}$ will affect the MLD performance. ${\lambda _L}( {{\bf{\tilde M}}} )$ can be expressed as
	\begin{equation}
		{\lambda _L}\left( {{\bf{\tilde M}}} \right) = {\lambda _L}\left( {{{\bf{P}}^H}{\bf{H}}^H{{\bf{H}}}{\bf{P}}} \right).
	\end{equation}
	To guarantee a promising MLD performance,  ${\bf{P}}$ must be a full-rank matrix, with the smallest singular value being maximized. This is due to the fact that a larger smallest singular value of ${\bf{P}}$ will yield a correspondingly larger smallest singular value of ${{\bf{\tilde M}}}$, thereby enhancing MLD performance. Based on the above analysis, our primary optimization objective for the subsequent problem is to maximize the smallest singular value of ${\bf{P}}$.
	
	\subsection{SLP Matrix Analysis}\label{ALPMA}
	In this section, we show analytically that the traditional SLP approach returns a rank-one precoding matrix, as stated in \textit{Corollary 1}.
	
	\textit{Corollary 1}: In P2, the optimized transmit SLP matrix ${\bf{P}}$ is rank-one.
	
	\textit{Proof}: We begin by transforming the power constraint in P2, where ${\bf{Ps}}$ can be decomposed as follows
	\begin{equation}
		{\bf{Ps}} = \sum\limits_{i = 1}^{KL} {{{\bf{p}}_i}{s_i}},\forall i \in \cal{KL}.
	\end{equation}
	${\bf{Ps}}$ can be viewed as a single vector variable for P2, and the distribution of power among each ${{{\bf{p}}_i}{s_i}}$ does not impact the optimal solution. Thus, each ${{{\bf{p}}_i}{s_i}}$ can be treated as identical, which results in
	\begin{equation}
	\left\| {{\bf{Ps}}} \right\|_2^2 = \left\| {KL{{\bf{p}}_i}{s_i}} \right\|_2^2 = {K^2}{L^2}s_i^*{\bf{p}}_i^H{{\bf{p}}_i}{s_i} = KL\sum\limits_{i = 1}^{KL} {s_i^*{\bf{p}}_i^H{{\bf{p}}_i}{s_i}},
	\end{equation}
	then the power constraint is equivalent to 
	\begin{equation}\label{Power}
	 \sum\limits_{i = 1}^{KL} {s_i^*{\bf{p}}_i^H{{\bf{p}}_i}{s_i}}  \le \frac{p}{{KL}}.	
	\end{equation}
	By replacing \textbf{C4} in P2 with (\ref{Power}), P2 can be further transformed into
	\begin{equation}
		\begin{array}{l}
			{\text{P6}}:\mathop {\min }\limits_{{\bf{P}},t,{\bf{\Gamma }}} {\kern 1pt} {\kern 1pt} {\kern 1pt} {\kern 1pt} -t \vspace{1ex}\\
			{\kern 23pt} s.t.{\kern 6pt} {\bf{C1}}:{{\bf{g}}_{k,l}}\sum\limits_{i = 1}^{KL} {{{\bf{p}}_i}{s_i}}  = {\bm{\gamma }}_{k,l}^T{{{\bf{\bar s}}}_{k,l}}, \forall k \in {\cal K},\forall l \in {\cal L} \vspace{1ex}\\
			{\kern 43pt} {\bf{C2}}:t - \gamma _m^{\cal O} \le 0,\forall \gamma _m^{\cal O} \in {\cal O} \vspace{1ex}\\
			{\kern 43pt} {\bf{C3}}:t - \gamma _n^{\cal I} = 0,\forall \gamma _n^{\cal I} \in {\cal I} \vspace{1ex}\\
			{\kern 43pt} {\bf{C4}}:\sum\limits_{i = 1}^{KL} {s_i^*{\bf{p}}_i^H{{\bf{p}}_i}{s_i}}  \le \dfrac{p}{{KL}},
		\end{array}
	\end{equation}
	where ${{\bf{g}}_{k,l}} = {{\bf{w}}_{k,l}}{{\bf{H}}_k}$. The Lagrangian function of P6 can be expressed as
	\begin{equation}\label{L1}
		\begin{array}{l}
			{\cal L}\left( {{{\bf{p}}_i},t,{\alpha _{k,l}},{\theta _m},{\vartheta _n},{\alpha _0}} \right) \vspace{1ex}\\
			  =- t + \sum\limits_{k = 1}^K {\sum\limits_{l = 1}^L {{\alpha _{k,l}}\left( {{{\bf{g}}_{k,l}}\sum\limits_{i = 1}^{KL} {{{\bf{p}}_i}{s_i}}  - {\bm{\gamma }}_{k,l}^T{{{\bf{\bar s}}}_{k,l}}} \right)} }  \vspace{1ex}\\
			{\kern 8pt}+ \sum\limits_{m = 1}^{card\left\{ O \right\}} {{\theta _m}\left( {t - \gamma _m^O} \right)}  + \sum\limits_{n = 1}^{card\left\{ I \right\}} {{\vartheta _n}\left( {t - \gamma _n^I} \right)} \vspace{1ex}\\
			 {\kern 8pt}+ {\alpha _0}\left( {\sum\limits_{i = 1}^{KL} {s_i^*{\bf{p}}_i^H{{\bf{p}}_i}{s_i}}  \le \dfrac{p}{{KL}}} \right),
		\end{array}
	\end{equation}
	where ${\alpha _{k,l}}$, ${\theta _m \ge 0}$, ${\vartheta _n}$ and ${\alpha _0 \ge 0}$ represent the introduced dual variables, and each ${\alpha _{k,l}}$ and ${\vartheta _n}$ can be complex. Furthermore, the KKT conditions of (\ref{L1}) can be derived as
	\begin{subequations}\label{KKT2}
		\begin{align}
			&\label{a1}\frac{{\partial {\cal L}}}{{\partial t}} =  - 1 + \sum\limits_{m = 1}^{card\left\{ O \right\}} {{\theta _m}}  + \sum\limits_{n = 1}^{card\left\{ I \right\}} {{\vartheta _n}}  = 0 \vspace{1ex}\\  
			&\label{a2}\frac{{\partial {\cal L}}}{{\partial {{\bf{p}}_i}}} = \left( {\sum\limits_{k = 1}^K {\sum\limits_{l = 1}^L {{\alpha _{k,l}}{{\bf{g}}_{k,l}}} } } \right){s_i} + {\alpha _0}{s_i}s_i^*{\bf{p}}_i^H = {\bf{0}},\forall i \in \cal{KL} \vspace{1ex}\\  
			&\label{a3}{{\bf{g}}_{k,l}}\sum\limits_{i = 1}^{KL} {{{\bf{p}}_i}{s_i}}  - {\bf{\gamma }}_{k,l}^T{{\bf{\bar s}}_{k,l}} = 0, \forall k \in {\cal K},\forall l \in {\cal L}\vspace{1ex}\\
			&\label{a4}{\theta _m}\left( {t - \gamma _m^O} \right) = 0,\forall \gamma _m^{\cal O} \in {\cal O}\vspace{1ex}\\
			&\label{a5}t - \gamma _n^{\cal I} = 0,\forall \gamma _n^{\cal I} \in {\cal I}\vspace{1ex}\\
			&\label{a6}{\alpha _0}\left( {\sum\limits_{i = 1}^{KL} {s_i^*{\bf{p}}_i^H{{\bf{p}}_i}{s_i}}  \le \frac{p}{{KL}}} \right) = 0.
		\end{align}
	\end{subequations}
	It can be observed that ${\alpha _0} \ne 0$ based on (\ref{a2}), and given ${\alpha _0 \ge 0}$, it follows that ${\alpha _0 > 0}$. This indicates the power constraint must be satisfied with equality when optimality is achieved. Therefore, ${\bf{p}}_i^H$ can be expressed as
	\begin{equation}\label{pH}
	{\bf{p}}_i^H =  - \frac{1}{{{\alpha _0}s_i^*}}\left( {\sum\limits_{k = 1}^K {\sum\limits_{l = 1}^L {{\alpha _{k,l}}{{\bf{g}}_{k,l}}} } } \right) = \frac{1}{{s_i^*}}\left( {\sum\limits_{k = 1}^K {\sum\limits_{l = 1}^L {{\chi _{k,l}}{{\bf{g}}_{k,l}}} } } \right),
	\end{equation}
	where we define
	\begin{equation}
	{\chi _{k,l}} =  - \frac{{{\alpha _{k,l}}}}{{{\alpha _0}}},\forall k \in {\cal K},\forall l \in {\cal L}.
	\end{equation}
	Based on (\ref{pH}), ${\bf{p}}_i$ can be expressed as
	\begin{equation}
	{{\bf{p}}_i} = \left( {\sum\limits_{k = 1}^K {\sum\limits_{l = 1}^L {\chi _{k,l}^*{\bf{g}}_{k,l}^H} } } \right)\frac{1}{{{s_i}}},\forall i \in \cal{KL},
	\end{equation}
	which further leads to
	\begin{equation}\label{ps}
	{{\bf{p}}_i}{s_i} = \sum\limits_{k = 1}^K {\sum\limits_{l = 1}^L {\chi _{k,l}^*{\bf{g}}_{k,l}^H} },\forall i \in \cal{KL}.
	\end{equation}
	It can be observed from (\ref{ps}) that ${{\bf{p}}_i}{s_i}$ is constant for any $i$. This indicates that each column of the precoding matrix satisfies: ${{\bf{p}}_i}= {{\bf{p}}_k} \cdot \dfrac{s_k}{s_i},\forall i,k$, which subsequently result in $rank({\bf P})=1$.
	
	Based on the above analysis, it is evident that the MLD method requires a larger smallest singular value of the transmit precoding matrix to enhance the decoding efficiency. However, the rank of the SLP matrix is proven to be one, which indicates that the MLD method cannot be directly used to decode the received signals transmitted by the SLP scheme. This limitation motivates us to propose more practical SLP-based schemes, which are tailored for MU-MIMO systems where users employ the MLD decoding approach.
	
	\subsection{Independent MLD Decoding}\label{IMLDDSLP}	
	In a typical MU-MIMO communication system, users independently decode their received signals using the MLD method. This is feasible because traditional BLP schemes are designed to eliminate or minimize inter-user interference, and the received signals contains only the desired signal and noise, where the inter-stream interference within the user itself is handled by the MLD procedure. Although the detection performance of the SLP method is expected to improve since the interference is pushed into the CI area in the traditional sense, it does not comply with the MLD rule because the received symbols are displaced from their nominal constellation points. This introduces additional challenges for independent MLD decoding for each user. To be more specific, we re-examine the received signal expression for both traditional BD precoding and SLP schemes. When the BS employs the traditional BD precoding scheme, the received signal for the \textit{k}-th user can be expressed as
	\begin{equation}\label{BLP}
		\begin{array}{l}
			{{\bf{y}}_k^{\tiny \text{BD}}} = {{\bf{H}}_k}\sum\limits_{k = 1}^K {{{\bf{P}}_k}{{\bf{s}}_k}}  + {{\bf{n}}_k} \vspace{1ex}\\
			{\kern 17pt}\mathop  = \limits^{\left( c \right)} {{\bf{H}}_k}{{\bf{P}}_k}{{\bf{s}}_k} + {{\bf{n}}_k} \vspace{1ex}\\
			{\kern 17pt}  = {{\bf{M}}_k}{{\bf{s}}_k} + {{\bf{n}}_k},
		\end{array}
	\end{equation}
	where step $(c)$ is achieved since the precoding matrix ${{\bf{P}}_k}$ for the \textit{k}-th user lies in the null space of the other users' channel matrices. (\ref{BLP}) has the same form with (\ref{estimation}) and ${{\bf{P}}_k}$ is a full-rank matrix, which indicates the MLD method can be readily applied. However, when the BS uses the SLP method, as shown in (\ref{P}) in this paper, the received signal for the \textit{k}-th user can be expressed as
	\begin{equation}
		\begin{array}{l}
			{\bf{y}}_k^{\tiny \text{SLP}} = {{\bf{H}}_k}\sum\limits_{k = 1}^K {{{\bf{P}}_k}{{\bf{s}}_k}}  + {{\bf{n}}_k}\vspace{1ex}\\
			{\kern 19pt} \mathop  = \limits^{\left( d \right)} K{{\bf{H}}_k}{{\bf{P}}_k}{{\bf{s}}_k} + {{\bf{n}}_k}\vspace{1ex}\\
			{\kern 19pt}  = K{{\bf{M}}_k}{{\bf{s}}_k} + {{\bf{n}}_k},
		\end{array}
	\end{equation}
	where step $(d)$ is implemented because ${{\bf{p}}_i}{s_i}$ is constant for any $i$ for QAM modulated symbols as proved above. Compared to the traditional BD precoding which fully eliminates inter-user interference for MU-MIMO, the received signal for the \textit{k}-th user increases by a factor of $K$ when using the SLP method. Although this result indicates that the SLP method enhances the users' SNR, the rank-one property of the SLP matrix renders the MLD method infeasible, as discussed in Section \ref{PAoMLD}. 
	
	\subsection{Proposed CSI-Free SLP-MLD Design}\label{transprecoding}	
	Based on the aforementioned analysis, we know that the smallest singular value of the SLP matrix should be maximized to enable MLD decoding in the users. However, in a typical MU-MIMO communication system, ${\bf{P}}$ is not a square matrix, which complicates the optimization of its smallest singular value.  Therefore, we propose to decompose ${\bf{P}}$ into a Hermitian matrix and a normal matrix. The lower bound of the smallest singular value of ${\bf{P}}$ can be obtained by maximizing the smallest singular value of the corresponding Hermitian matrix. Consequently, a full-rank SLP matrix can be derived and the MLD method can be applied to decode the received signals.
	
	Specifically, given that ${\bf{P}}$ is a ${{N_T \times KL}}$ dimensional matrix and ${N_t} \ge KL$ for MU-MIMO systems, ${\bf{P}}$ can be decomposed into two parts as
	\begin{equation}
		{\bf{P}} = \left[ {\begin{array}{*{20}{c}}
				{{{\bf{P}}_1}}\\
				{{{\bf{P}}_2}}
		\end{array}} \right],
	\end{equation}
	where ${\bf{P}}_1 \in \mathbb{C} {^{KL \times KL}}$ is assumed to be a Hermitian matrix and ${\bf{P}}_2$ is a normal matrix of dimensions $\left( N_T - KL \right) \times KL$. Based on the aforementioned analysis, ${\bf{P}}$ will be full-rank if ${\bf{P}}_1$ is optimized to be a full-rank matrix. Furthermore, according to the fundamental rank properties\cite{MatrixAnalysis}, it can be established that
	\begin{equation}
		{\sigma _{\min }}\left( {\bf{P}} \right) \ge {\sigma _{\min }}\left( {{{\bf{P}}_1}} \right) = {\lambda _{\min }}\left( {{{\bf{P}}_1}} \right),
	\end{equation}
	where ${\sigma _{\min }}\left( {\bf{P}} \right)$ represents the smallest singular value of ${\bf{P}}$. It can be observed that the smallest eigenvalue of ${\bf{P}}_1$ is the lower bound of the smallest singular value of ${\bf{P}}$. Therefore, the task of maximizing the smallest singular value of  ${\bf{P}}$ can be transformed into maximizing the smallest eigenvalue of ${\bf{P}}_1$. Thus, the smallest singular value maximization problem can be expressed as
	\begin{equation}\label{P9} 
		\begin{array}{l}
			{\text{P7}}:\mathop {\max }\limits_{{{\bf{P}}_1},{{\bf{P}}_2},t,{{\bf{\Gamma }}}} {\kern 4pt}{\lambda _{\min }}\left( {{{\bf{P}}_1}} \right) \vspace{1ex}\\
			{\kern 31pt} s.t.{\kern 15pt}{\bf{C1}}:{\bf{P}} = \left[ {\begin{array}{*{20}{c}}
					{{{\bf{P}}_1}}\\
					{{{\bf{P}}_2}}
			\end{array}} \right],{{\bf{P}}_1}{\kern 1pt} {\kern 1pt} {\rm{is}}{\kern 1pt} {\kern 1pt} {\rm{Hermitian}}{\kern 1pt} {\kern 1pt} {\rm{matrix}} \vspace{1ex}\\
			{\kern 60pt} {\bf{C2}}:{\bf{GPs}} = {\bf{U}}\text{diag}\left( {\bf{\Gamma }} \right){\bf{\bar s}} \vspace{1ex}\\
			{\kern 60pt}{\bf{C3}}:t - \gamma _{m}^{\cal O} \le 0,\forall \gamma _{m}^{\cal O} \in {\cal O} \vspace{1ex}\\
			{\kern 60pt}{\bf{C4}}:t - \gamma _{n}^{\cal I} = 0,\forall \gamma _{n}^{\cal I} \in {\cal I} \vspace{1ex}\\
			{\kern 60pt}{\bf{C5}}:\left\| {{\bf{Ps}}} \right\|_2^2 \le p \vspace{1ex}\\
			{\kern 60pt}{\bf{C6}}:{\kern 1pt} {\kern 1pt} {\kern 1pt} {\bf{Ps}} = K{{\bf{P}}_k}{{\bf{s}}_k},\forall k \in {\cal K},
		\end{array}
	\end{equation}
	where $\bf{C1}$ ensures that ${\bf{P}}_1$ is a Hermitian matrix and $\bf{C6}$ is the independent MLD decoding requirement. P7 is a convex optimization problem and can be directly solved by using the CVX tool.

	Furthermore, the analysis of the MLD method in Section \ref{PAoMLD} reveals that a larger smallest singular value of the transmit precoding matrix is more beneficial to the decoding efficiency of the users. Therefore, the CI constraints $\bf{C2}$-$\bf{C4}$, which aim to transform multi-user interference into CI are no longer necessary for optimizing the singular values, as users utilize the MLD method to decode received signals. Thus, we can reformulate P7 into
	\begin{equation}
		\begin{array}{l}
			{\text{P8}}:\mathop {\max }\limits_{{{\bf{P}}_1},{{\bf{P}}_2},z} {\kern 4pt} z \vspace{1ex}\\
			{\kern 33pt}  s.t. {\kern 12pt} {\bf{C1}}:\left\| {{{\bf{P}}_1}{\bf{s}}} \right\|_2^2 + \left\| {{{\bf{P}}_2}{\bf{s}}} \right\|_2^2 \le p \vspace{1ex}\\
			{\kern 59pt} {\bf{C2}}:{{\bf{P}}_1}{\bf{s}} = K{{\bf{P}}_{k,1}}{{\bf{s}}_k},\vspace{1ex}\\
			{\kern 82pt}{{\bf{P}}_2}{\bf{s}} = K{{\bf{P}}_{k,2}}{{\bf{s}}_k},\forall k \in {\cal K} \vspace{1ex}\\
			{\kern 59pt}{\bf{C3}}:\left[ {\begin{array}{*{20}{c}}
					{{\cal R}\left( {{{\bf{P}}_1}} \right)}&{ - {\cal I}\left( {{{\bf{P}}_1}} \right)} \vspace{1ex}\\
					{{\cal I}\left( {{{\bf{P}}_1}} \right)}&{{\cal R}\left( {{{\bf{P}}_1}} \right)}
			\end{array}} \right] - z{\bf{I}} \succeq {\bf{0}},
		\end{array}
	\end{equation}
	where ${{\bf{P}}_{k,1}}$ is a square matrix composed of the first $L$ rows of ${{\bf{P}}_{k}}$, and ${{\bf{P}}_{k,2}}$ includes the last ${N_T}-L$ rows of ${{\bf{P}}_{k}}$, and the objective function $z$ represents the optimized smallest singular value of ${{\bf{P}}_1}$. It can be observed that only the transmit symbols are utilized to design the SLP matrix, while the CSI matrices are omitted. This approach not only simplifies the optimization problem but also mitigates the performance losses associated with inaccurate channel estimation.
	
	Notably, our goal is to maximize the smallest singular value of ${{\bf{P}}_1}$, which is uncorrelated with ${{\bf{P}}_2}$. Therefore, more power should be allocated to ${{\bf{P}}_1}$ and the variables can be decoupled by decomposing P8 into two subproblems as
	\begin{equation}
		\begin{array}{l}
			{\text{P9.1}}: {\kern 4pt} \mathop {\min }\limits_{{{\bf{P}}_2}}  {\kern 3pt} \left\| {{{\bf{P}}_2}{\bf{s}}} \right\|^2_2 \vspace{1ex}\\
			{\kern 35pt} s.t.{\kern 8pt}{\bf{C1}}:{{\bf{P}}_2}{\bf{s}} = K{{\bf{P}}_{2,k}}{{\bf{s}}_k},\forall k \in {\cal K},
		\end{array}
	\end{equation}
	and 
	\begin{equation}
		\begin{array}{l}
			{\text{P9.2}}:{\kern 4pt} \mathop {\max }\limits_{{{\bf{P}}_1},z} {\kern 5pt} z \vspace{1ex}\\
			{\kern 37pt} s.t. {\kern 7pt}{\bf{C1}}: \left\| {{{\bf{P}}_1}{\bf{s}}} \right\|_2^2 \le p \vspace{1ex}\\
			{\kern 58pt}{\bf{C2}}:{{\bf{P}}_1}{\bf{s}} = K{{\bf{P}}_{1,k}}{{\bf{s}}_k},\forall k \in {\cal K} \vspace{1ex}\\
			{\kern 58pt}{\bf{C3}}:\left[ {\begin{array}{*{20}{c}}
					{{\cal R}\left( {{{\bf{P}}_1}} \right)}&{ - {\cal I}\left( {{{\bf{P}}_1}} \right)} \vspace{1ex}\\
					{{\cal I}\left( {{{\bf{P}}_1}} \right)}&{{\cal R}\left( {{{\bf{P}}_1}} \right)}
			\end{array}} \right] - z{\bf{I}} \succeq {\bf{0}}.
		\end{array}
	\end{equation}
	It is evident that the optimal solution to P9.1 is ${\bf{P}}_2^* = {\bf{0}}$, which further indicates that only ${{\bf{P}}_1}$ needs to be optimized. Moreover, P9.2 is a simpler convex problem with fewer variables, and can be solved by using the CVX tool.
	
	Although P9.2 includes fewer variables that need to be optimized compared to P7, which simplifies the problem, this advantage may result in a degradation of communication performance. This degradation primarily arises from the omission of the CI constraints. First, these constraints convert interference into useful signals, thereby enhancing the receiver's decoding capability. Second, they ensure that ${{\bf{P}}_2}$ is a non-zero matrix, which can potentially increase the singular values of the precoding matrix ${{\bf{P}}}$. 
	
	\section{Convergence and Computational Complexity Analysis} \label{CCCA}
	In this section, we analyze the convergence of the proposed AO algorithm for the joint design scheme. Subsequently, we derive the computational complexity of the proposed joint design scheme, the SSVMP scheme, the SDP-based scheme and the MLD method.
	
	\subsection{Convergence Analysis}\label{CA}
	The solution to our proposed joint design scheme is attained through the AO method, and the convergence of this algorithm is ensured for two key reasons: 1) After each iteration, the value of the objective function is monotonically increasing. This characteristic guarantees that the algorithm consistently progresses toward an improved solution with each iteration \cite{PSK SLP MU-MIMO}. 2) The proposed problem has an upper bound on the objective function value, determined by the CI constraints. The above two observations guarantee convergence, ensuring that the optimization process remains stable and does not exceed feasible limits \cite{ConvexOptimization}, which is also validated by our numerical results to be shown in Section \ref{Simulation}.
	
	\subsection{Computational Complexity Analysis}\label{CCA}
	The overall complexity of the proposed \textbf{Algorithm \ref{al1}} can be divided into two components: the optimization of the receive combining matrix ${\bf{W}}_k$ and the computation of the precoding matrix ${\bf{P}}$. Both components involve solving convex problems, which ensures that optimal solutions can be achieved. First, the iterative algorithm introduced in \cite{CF-QAM} effectively solves the two QP subproblems to obtain the dual variables ${\bf{u}}_1$ and ${{\bf{u}}}$. Both QP problems have the same number of variables, and the primary computational complexity arises from matrix inversion, leading to a computational complexity of ${\mathcal{O}}({(2KL)}^{3})$ for each subproblem. Second, the most computationally expensive operation within the calculation of the two closed-form solutions is also the matrix inversion, which has a complexity of approximately ${\mathcal{O}}({8K^{3}L^{3}})$. Consequently, if we denote $N_{iter}$ as the total number of iterations, the overall computational complexity of \textbf{Algorithm \ref{al1}} can be expressed as $N_{iter}(2{\mathcal{O}}({(2KL)}^{3})+{\mathcal{O}}({8K^{3}L^{3}}))$ \cite{PSK SLP MU-MIMO}.
	
	Besides using the receive combining matrix, we propose to utilize the MLD method to decode the received signals at the receivers. The computational complexity of the MLD method is $K{L^{M_c}}$, where ${M_c}$ represents the modulation order of QAM constellation , and each user independently decodes the data symbols. It is evident that the computational complexity increases exponentially, leading to a significant computational burden in MU-MIMO systems. To address this issue, the QRM-MLD method is considered as an alternative, with a computational complexity of $K\left[ {{M_c} + M\left( {L - 1} \right)} \right]$, where $M$ denotes the number of surviving symbol candidates that must be calculated starting from the second stage. Clearly, QRM-MLD method significantly reduces the computational complexity.
	
%	\begin{table}[!b]
%		\renewcommand{\arraystretch}{1.5}
%		\centering
%		\caption{Computation complexity of considered schemes}
%		\begin{tabular}{ |c|c|c|c|c| } 
%			\hline
%			Scheme & Computation Complexity  \\
%			\hline
%			\textbf{Algorithm \ref{al1}}& $N_{iter}(2{\mathcal{O}}({(2KL)}^{3})+{\mathcal{O}}({8K^{3}L^{3}}))$ \\ 
%			\hline
%			\text{SSVMP} &${\mathcal{O}}({({N_T}KL)}^{3.5})$ \\
%			\hline
%			SDP-based & ${\mathcal{O}}({(KL)}^{7})$\\
%			\hline
%			MLD & $K{L^{M_c}}$\\
%			\hline
%			QRM-MLD & $K\left[ {{M_c} + M\left( {L - 1} \right)} \right]$\\
%			\hline
%		\end{tabular}
%		\label{table}
%	\end{table} 
	
	Additionally, two novel symbol-level singular value optimization problems are introduced to enhance the decoding efficiency. SSVMP in (\ref{P9}) is a convex matrix block optimization problem that can be directly solved by using the CVX tool, resulting in a computational complexity of ${\mathcal{O}}({({N_T}KL)}^{3.5})$, where the number of variables is ${N_T}KL$. To further reduce computational complexity, an SDP-based problem is introduced, which requires only the calculation of ${\bf{P}}_1$. The computational complexity is ${\mathcal{O}}({(KL)}^{7})$, with the number of variables reduced to ${K^2L^2}$.

	\section{Simulation Results}\label{Simulation}	
	
	\begin{figure}[!b]
		\centering
		\includegraphics[width=3.5in]{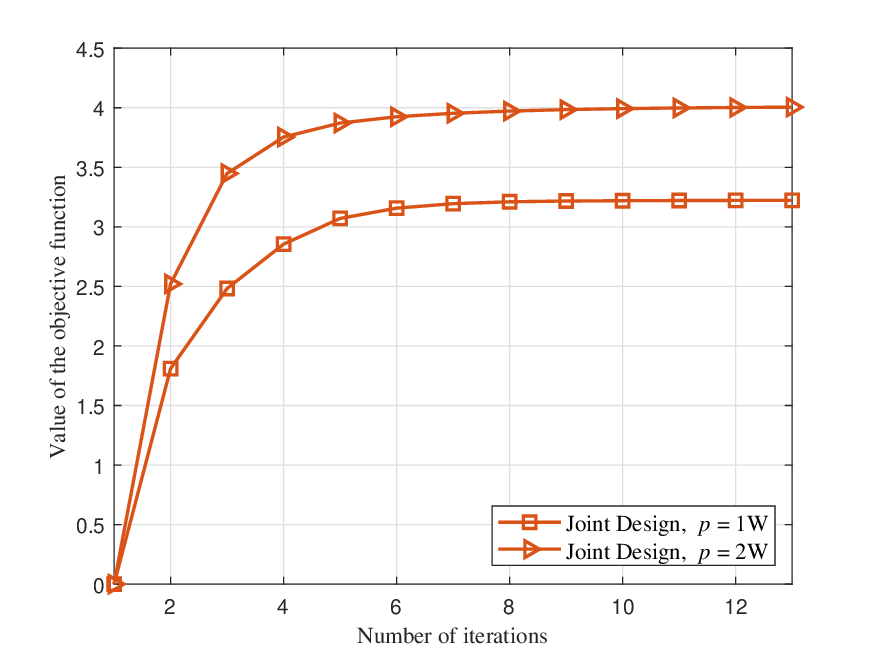}
		\caption{Convergence for the proposed Joint Design scheme, 16QAM, $N_T=16, N_R=8, L=4$ and $K=2$.}
		\label{convergence}
	\end{figure}
	
	\begin{figure*}[!t]
		\centering
		\begin{minipage}[b]{0.45\linewidth}
			\centering
			\includegraphics[width=\linewidth]{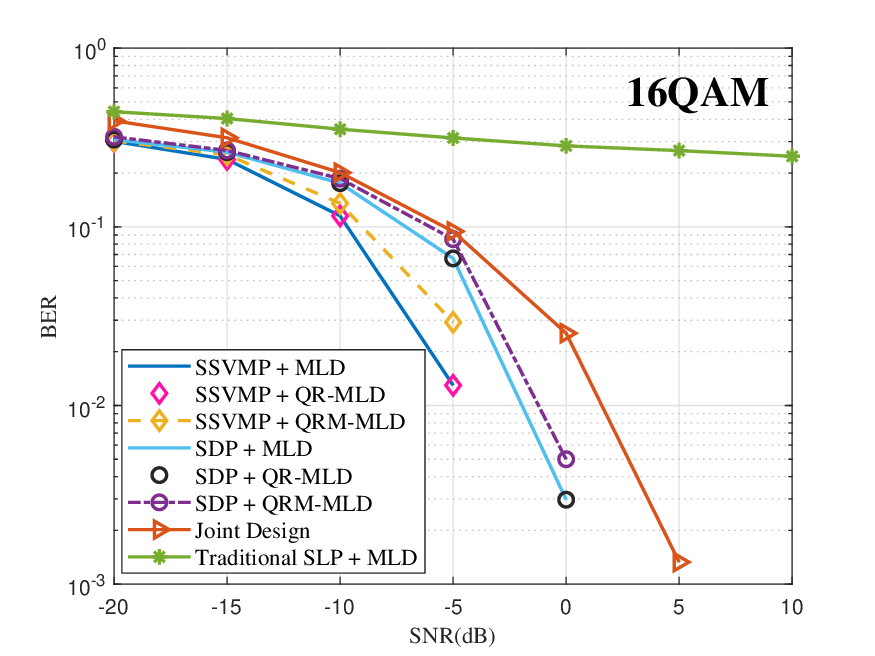} % 修改为新的文件名
			\caption{BER v.s. SNR, 16QAM, ${N_T}=16, {N_R}=8, L=4$, $ K=2$ and $M=8$.}
			\label{16QAM_MLD_QRMMLD} % 修改为新的label
		\end{minipage}
		\hfill
		\begin{minipage}[b]{0.45\linewidth}
			\centering
			\includegraphics[width=\linewidth]{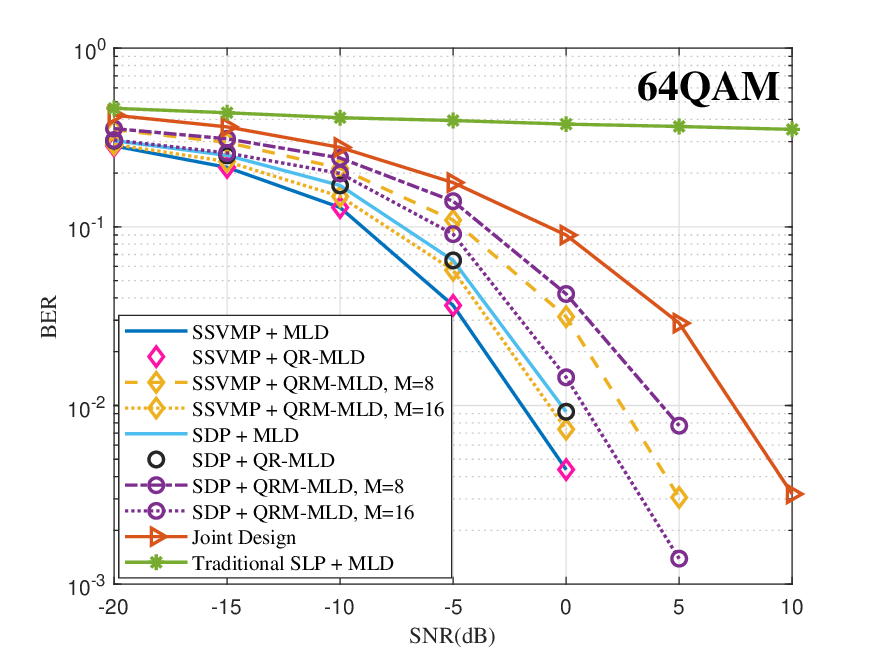} % 修改为新的文件名
			\caption{BER v.s. SNR, 64QAM, ${N_T}=16, {N_R}=8, L=4$, $ K=2$ and $M=8 \ or \ 16$.}
			\label{64QAM_MLD_QRMMLD} % 修改为新的label
		\end{minipage}
	\end{figure*}
	
	\begin{figure*}[!t]
		\centering
		\begin{minipage}[b]{0.45\linewidth}
			\centering
			\includegraphics[width=\linewidth]{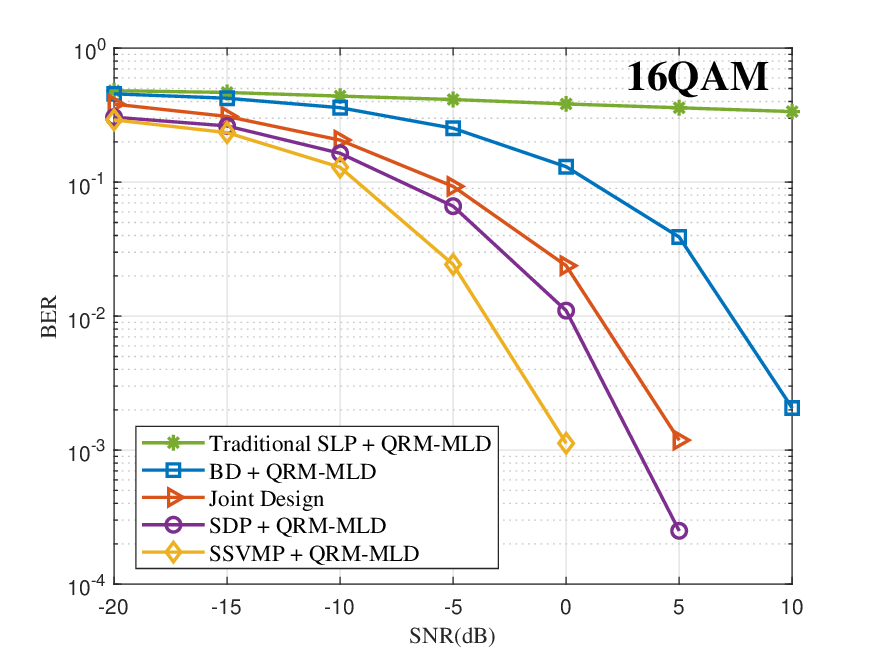} % 修改为新的文件名
			\caption{BER v.s. SNR, 16QAM, ${N_T}=32, {N_R}=8, L=4, K=2$ and $ M=8$.}
			\label{32-8-4-2} % 修改为新的label
		\end{minipage}
		\hfill
		\begin{minipage}[b]{0.45\linewidth}
			\centering
			\includegraphics[width=\linewidth]{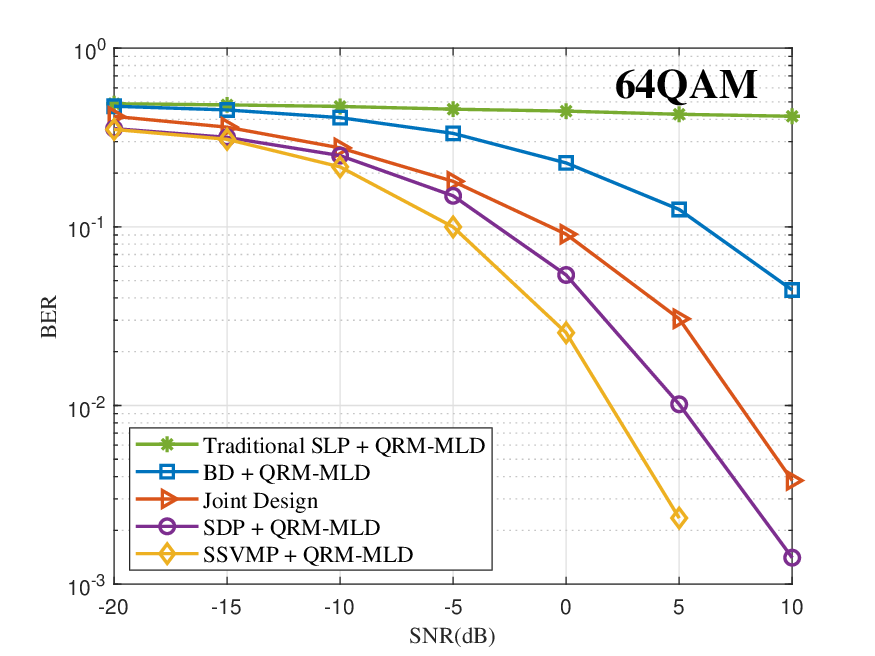} % 修改为新的文件名
			\caption{BER v.s. SNR, 64QAM, ${N_T}=32, {N_R}=8, L=4, K=2$ and $ M=8$.}
			\label{(64QAM)32-8-4-2} % 修改为新的label
		\end{minipage}
	\end{figure*}
	In this section, we provide a comparative analysis of the numerical results for the proposed joint design scheme alongside two singular value optimization problems, in contrast to the traditional SLP and BD precoding schemes paired with MLD estimation at the receivers. MonteCarlo simulations are utilized as the evaluation methodology. For each time slot, the transmit power budget is set to ${P_T}=1\,\text{W}$, and the transmit SNR is defined as $\rho=1/{\sigma^2}$. The elements of the channel matrix ${\bf{H}}$ are assumed to follow a standard complex Gaussian distribution, specifically ${\bf{H}}_{m,n} \sim \mathcal{CN}(0,1)$. To ensure clarity, the following abbreviations will be consistently used throughout this section:
	
	\begin{enumerate}
		\item `Traditional SLP': traditional optimization-based SLP scheme based on P2.
		\item `BD': traditional BD scheme as proposed in \cite{BD};
		\item `Joint Design': proposed joint symbol-level transmit precoding and receive combining scheme based on \textbf{Algorithm \ref{al1}};
		\item `SSVMP': proposed smallest singular value maximization problem (SSVMP) based on P7;
		\item `SDP': proposed SDP-based singular value optimization problem based on P9.2;
		\item `MLD', `QR-MLD' or `QRM-MLD': the MLD, QR-MLD or QRM-MLD methods used at the user side to decode the receive signals.
	\end{enumerate} 
	
	Fig. \ref{convergence} illustrates the convergence of the `Joint Design' scheme proposed in \textbf{Algorithm \ref{al1}}, employing 16QAM constellation with $N_T=16, N_R=8, L=4$ and $K=2$. In each iteration, the optimized objective $t$ between the current and the previous iteration for different transmit power settings is distinguished, and it increases with the transmit power. This is because more power can be used to increase the distance between the detection threshold and the CI region. We also observe that the proposed algorithm converges only within several iterations. This finding is consistent with the convergence analysis presented in Section \ref{CA}.	
	
	In Fig. \ref{16QAM_MLD_QRMMLD} and Fig. \ref{64QAM_MLD_QRMMLD}, we compare the BER performance of the proposed against the `Traditional SLP' scheme when using various decoding techniques for 16QAM and 64QAM constellations, with parameters set to $N_T=16, N_R=8, L=4$ and $K=2$. It can be observed that the BER performance of the `Traditional SLP + MLD' scheme experiences significant degradation, which validates the analysis proposed in Section \ref{PAoMLD} and Section \ref{ALPMA} and highlights the necessity of our work. Furthermore, our proposed schemes achieves promising BER performance, this is attributed to the utilization of symbol-by-symbol optimization and the MLD method. In both figures, the `SSVMP' scheme achieves the best performance, and the `SDP' scheme incurs only an acceptable performance loss compared to the `SSVMP' scheme. In addition, the `QR-MLD' method can achieve the same performance with `MLD' method with lower computational complexity. To further reduce the computational complexity, the QRM-MLD method is a more practical and efficient alternative. Therefore, we evaluate the BER performance of the `QRM-MLD' method for clarity, setting $M=8$ and $M=16$. Notably, when $M=16$, the `MLD' method is equivalent to the `QRM-MLD' method for 16QAM constellation. It can be observed that the `QRM-MLD' method exhibits a slight performance loss compared to the `MLD' method, and a larger value of $M$ leads to better performance, this indicates there is a trade-off between the performance and complexity. To further verify the effectiveness of the proposed schemes, subsequent simulations will directly employ the `QRM-MLD' method. Additionally, Fig. \ref{16QAM_MLD_QRMMLD} achieves a lower BER compared to Fig. \ref{64QAM_MLD_QRMMLD}, this is due to the smaller distance between symbols and there are more inner points at higher modulation orders, which increases the likelihood of decoding errors.
	
	\begin{figure}[!t]
		\centering
		\includegraphics[width=3.5in]{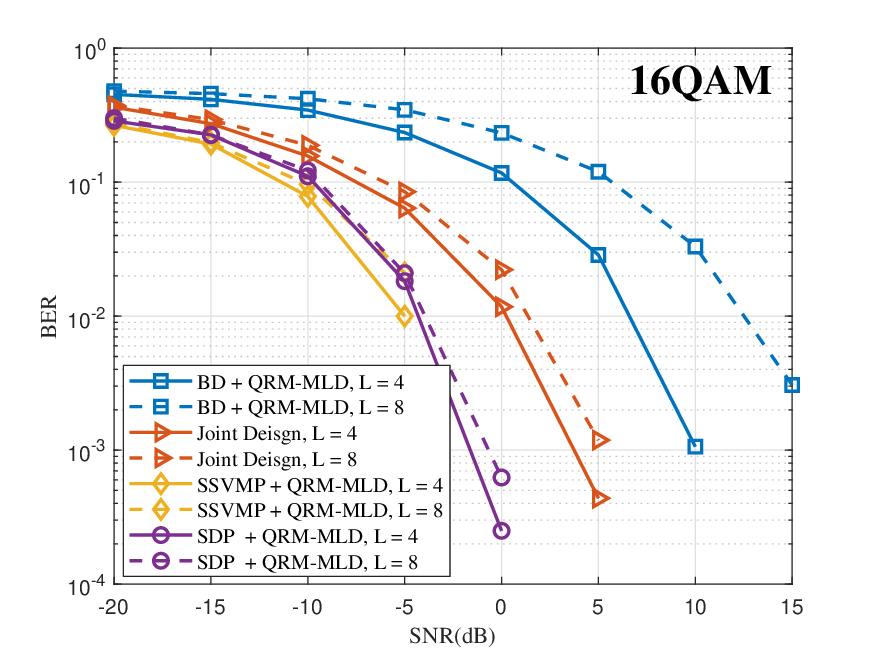}
		\caption{BER v.s. SNR for different number of data streams per user, 16QAM, ${N_T}=32, {N_R}=16$ and $ K=2$.}
		\label{datastreams}
	\end{figure}
	\begin{figure}[!t]
		\centering
		\includegraphics[width=3.5in]{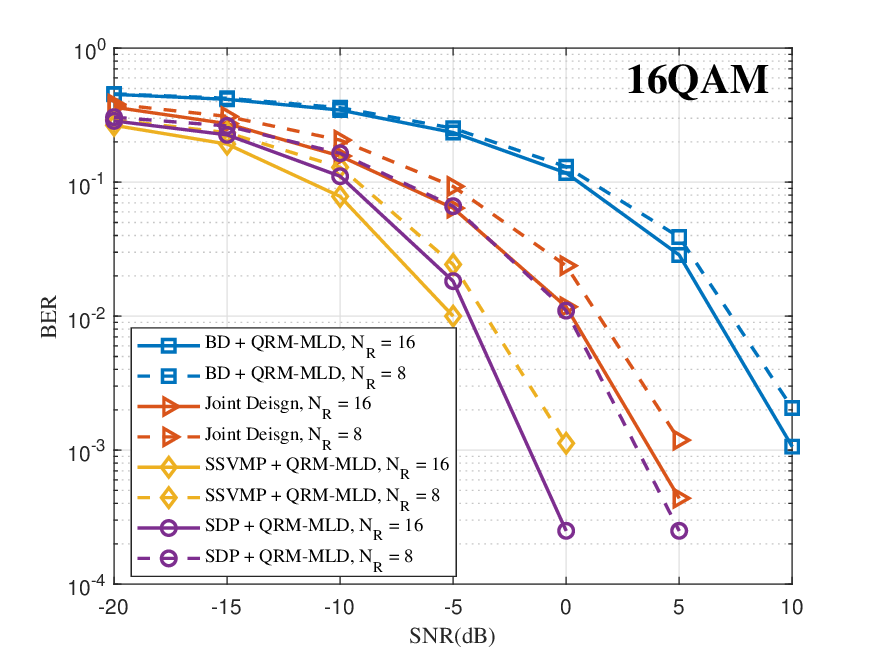}
		\caption{BER v.s. SNR for different number of receive antennas, 16QAM, ${N_T}=32, L=4$ and $ K=2$.}
		\label{receiveantennas}
	\end{figure}
	\begin{figure}[!t]
		\centering
		\includegraphics[width=3.5in]{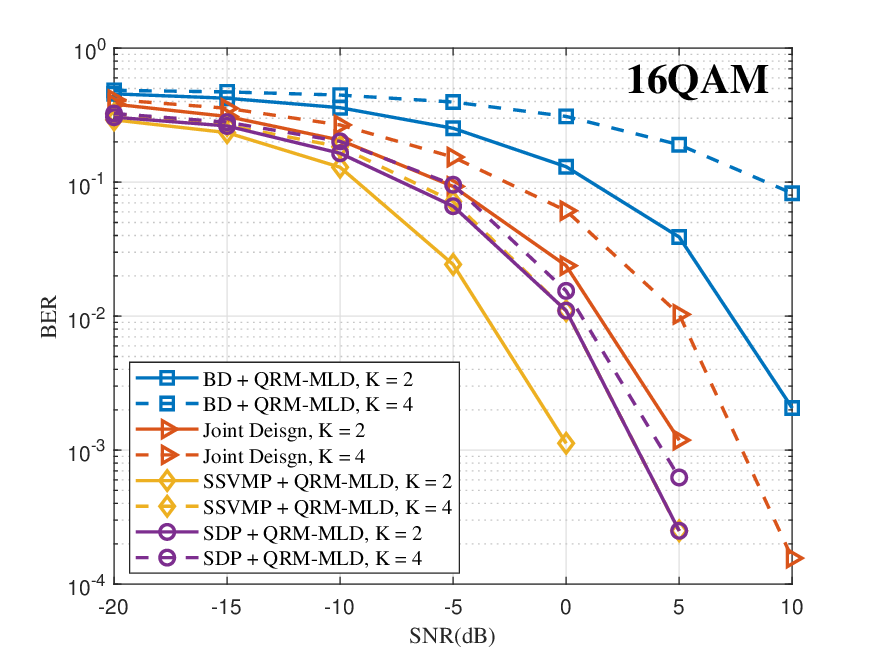}
		\caption{BER v.s. SNR for different number of users, 16QAM, ${N_T}=32, {N_R}=8$ and $ L=4$.}
		\label{users}
	\end{figure}

	In Fig. \ref{32-8-4-2} and Fig. \ref{(64QAM)32-8-4-2}, we compare the BER performance of our proposed schemes to `Traditional SLP' and `BD' schemes employing 16QAM and 64QAM constellations. In both figures, $N_T=32, {N_R}=8, L=4,  K=2$ and $M=8$. It can be observed that the BER decreases with the increase of the SNR, and higher BER gains can be achieved in the high SNR regime. It's evident that the proposed SLP schemes significantly outperform the `Traditional SLP' and `BD' schemes, and the performance of the `Traditional SLP + QRM-MLD' completely fails, which indicates that traditional SLP approach is not compatible with MLD method. Moreover, the performance improvement is more pronounced in the case of `SSVMP' scheme and `SDP' scheme compared to the `Joint Design' scheme, this benefits from the SLP design tailored specifically for MLD method. In addition, we observe that Fig. \ref{32-8-4-2} achieves a lower BER compared to Fig. \ref{(64QAM)32-8-4-2} under the same SNR conditions, since the lower modulation order symbols are transmitted.
	
	Fig. \ref{datastreams} shows the BER performance with respect to the number of data streams per user employing 16QAM constellation. In the figure, the solid and the dashed lines represent $L=4$ and $L=8$, respectively, with ${N_T}=32, {N_R}=16, K=2$. Similar BER performance is observed across the different schemes. However, the BER increases as the number of data streams grows under the same schemes, this can be attributed to the reduction in design flexibility resulting from the increased number of data streams. Furthermore, it can be observed that when the number of data streams is relatively large, the performance gap between the `SSVMP ' scheme and the `SDP' scheme diminishes. This observation renders the low-complexity `SDP' scheme more practical.
	
	Fig. \ref{receiveantennas} depicts the BER performance of various precoding approaches with respect to the number of the receive antennas while using 16QAM constellation, with ${N_T}=32, L=4, K=2$. In the figure, the solid and the dashed lines represent $N_R=16$ and $N_R=8$, respectively. It can be observed that BER decreases with the increase of the number of the receive antennas. This is because when users are equipped with more receive antennas, the rank of the channel matrix ${\bf{H}}_{k}$ increases, which improves the estimation performance of the MLD method. In addition, more design freedoms are available as the number of receive antennas increases, which offers more effective design for `Joint Design' scheme.
	
	Fig. \ref{users} illustrates the BER performance concerning the number of users, employing 16QAM constellation with ${N_T}=32, {N_R}=8, L=4$. In the figure, the solid and the dashed lines represent $K=2$ and $K=4$, respectively. It's evident that the BER increases with the increase of the number of users. This phenomenon is attributed to the fact that the increasing number of users results in fewer design freedoms available, until $N_T = KL$. At this point, the one-to-one mapping between transmit antennas and data streams imposes performance limitations.

	\section{Conclusion}\label{Conclusion}
	In this paper, the symbol-level transmit precoding and receive decoding techniques are studied for MU-MIMO communication systems while employing QAM constellation symbols. A joint transmit precoding and receive combining optimization problem is first formulated, and the optimal structures are derived by analyzing the Lagrangian and KKT conditions. Additionally, the MLD method is used to decode the received signals, addressing the dependence of the receive combining matrix on the transmit symbols. Furthermore, to overcome the rank-deficient problem of the SLP matrix, two novel symbol-level singular value optimization problems are proposed to enhance the decoding efficiency. Numerical results show the superior performance improvement of the proposed schemes over the traditional BD-based approach.
	
	\appendices
	\section{Proof for Proposition 1}
	We first use the row vectors ${{{\bf{w}}_{k,l}}}$ to replace the original matrix variables ${{{\bf{W}}_{k}}}$, then P4 can be transformed into 
	\begin{equation}\label{w}
		\begin{array}{l}
			{\text{P10}}:\mathop {\min }\limits_{{{{\bf{w}}_{k,l}}},t,{\bm{\Gamma }}} {\kern 4pt}  - t  \vspace{1ex}\\
			{\kern 33pt} s.t.{\kern 15pt} {\bf{C1}}:{{\bf{w}}_{k,l}}{{\bf{r}}_k} = {\bm{\gamma }}_{k,l}^T{{{\bf{\bar s}}}_{k,l}},{\kern 1pt} \forall k \in {\cal K},\forall l \in {\cal L}  \vspace{1ex}\\
			{\kern 62pt} {\bf{C2}}:t - \gamma _m^{\cal O} \le 0,\forall \gamma _m^{\cal O} \in {\cal O}  \vspace{1ex}\\
			{\kern 62pt} {\bf{C3}}:t - \gamma _n^{\cal I} = 0,\forall \gamma _n^{\cal I} \in {\cal I}  \vspace{1ex}\\
			{\kern 62pt} {\bf{C4}}:\sum\limits_{l = 1}^L {{{\bf{w}}_{k,l}}{\bf{w}}_{k,l}^H}  \le 1,{\kern 1pt} \forall k \in {\cal K},
		\end{array}
	\end{equation}
	where ${{{\bf{w}}_{k,l}}}$ is the \textit{l}-th row of ${{{\bf{W}}_{k}}}$. The Lagrangian function of P10 can be expressed as
	\begin{equation}\label{Lw}
	\begin{array}{l}
		{\cal L}\left( {{{\bf{w}}_{k,l}},t,{\delta _{k,l}},{\mu _m},{\nu _n},{\varphi _k}} \right) \vspace{1ex}\\
		=  - t + \sum\limits_{k = 1}^K {\sum\limits_{l = 1}^L {{\delta _{k,l}}\left( {{{\bf{w}}_{k,l}}{{\bf{r}}_k} - {\bm{\gamma }}_{k,l}^T{{{\bf{\bar s}}}_{k,l}}} \right)} } \vspace{1ex}\\
		 {\kern 10pt} + \sum\limits_{m = 1}^{card\left\{ {\cal O} \right\}} {{\mu _m}\left( {t - \gamma _m^{\cal O}} \right)} +
		 \sum\limits_{n = 1}^{card\left\{ {\cal I} \right\}} {{\nu _n}\left( {t - \gamma _n^{\cal I}} \right)} \vspace{1ex}\\
		 {\kern 10pt}+ \sum\limits_{k = 1}^K {{\varphi _k}\left( {\sum\limits_{l = 1}^L {{{\bf{w}}_{k,l}}{\bf{w}}_{k,l}^H}  - 1} \right)} ,
	\end{array}
	\end{equation}
	where ${\delta _{k,l}}$, ${\mu _m \ge 0}$, ${\nu _n}$ and ${\varphi _k \ge 0}$ are the dual variables. Each ${\delta _{k,l}}$ and ${\nu _n}$ may be complex. Based on the Lagrangian in (\ref{Lw}), the KKT conditions for optimality can be obtained as
	\begin{subequations}\label{KKT1}
		\begin{align}
			&\label{c1}\frac{{\partial {\cal L}}}{{\partial t}} =  - 1 + \sum\limits_{m = 1}^{card\left\{ {\cal O} \right\}} {{\mu _m}}  + \sum\limits_{n = 1}^{card\left\{ {\cal I} \right\}} {{\nu _n}}  = 0 \vspace{1ex}\\  
			&\label{c2}\frac{{\partial {\cal L}}}{{\partial {{\bf{w}}_{k,l}}}} = {\delta _{k,l}}{{\bf{r}}_k} + {\varphi _k}{\bf{w}}_{k,l}^H = {\bf{0}},{\kern 1pt} \forall k \in {\cal K},\forall l \in {\cal L} \vspace{1ex}\\  
			&\label{c3}{{\bf{w}}_{k,l}}{{\bf{r}}_k} - {\bf{\gamma }}_{k,l}^T{{\bf{\bar s}}_{k,l}} = 0,{\kern 1pt} \forall k \in {\cal K},\forall l \in {\cal L} \vspace{1ex}\\
			&\label{c4}{\mu _m}\left( {t - \gamma _m^{\cal O}} \right) = 0,\forall \gamma _m^{\cal O} \in {\cal O}\vspace{1ex}\\
			&\label{c5}t - \gamma _n^{\cal I} = 0,\forall \gamma _n^{\cal I} \in {\cal I}\vspace{1ex}\\
			&\label{c6}{\varphi _k}\left( {\sum\limits_{l = 1}^L {{{\bf{w}}_{k,l}}{\bf{w}}_{k,l}^H}  - 1} \right) = 0.
		\end{align}
	\end{subequations}
	
	In order to satisfy conditions (\ref{c2}) and ${\varphi _k \ge 0}$, it can be concluded that ${\varphi _k > 0}$. Therefore, ${{{\bf{w}}_{k,l}}}$ can be derived as
	\begin{equation}
	\begin{array}{l}
		{\kern 13pt}{\bf{w}}_{k,l}^H =- \dfrac{{{\delta _{k,l}}}}{{{\varphi _k}}}{{\bf{r}}_k} = {\zeta _{k,l}}{{\bf{r}}_k} \vspace{1ex}\\
		\Rightarrow {{\bf{w}}_{k,l}} = \zeta _{k,l}^*{\bf{r}}_k^H,{\kern 1pt} \forall k \in {\cal K},\forall l \in {\cal L},
	\end{array}
	\end{equation}
	where we define ${\zeta _{k,l}} \buildrel \Delta \over =- \dfrac{{{\delta _{k,l}}}}{{{\varphi _k}}}$. The receive combining matrix can be further expressed as
	\begin{equation}
	{{\bf{W}}_k} = \left[ {\begin{array}{*{20}{c}}
			{{{\bf{w}}_{k,1}}}\\
			{{{\bf{w}}_{k,2}}}\\
			\vdots \\
			{{{\bf{w}}_{k,L}}}
	\end{array}} \right] = \left[ {\begin{array}{*{20}{c}}
			{\zeta _{k,1}^*{\bf{r}}_k^H}\\
			{\zeta _{k,2}^*{\bf{r}}_k^H}\\
			\vdots \\
			{\zeta _{k,L}^*{\bf{r}}_k^H}
	\end{array}} \right] = \left[ {\begin{array}{*{20}{c}}
			{\zeta _{k,1}^*}\\
			{\zeta _{k,2}^*}\\
			\vdots \\
			{\zeta _{k,L}^*}
	\end{array}} \right]{\bf{r}}_k^H = {\bm{\zeta }}_k^*{\bf{r}}_k^H,
	\end{equation}
	where ${{\bm{\zeta }}_k} = {\left[ {{{\bf{\zeta }}_{k,1}},{{\bf{\zeta }}_{k,2}}, \cdots ,{{\bf{\zeta }}_{k,L}}} \right]^T}$. According to constraint \textbf{C1} in P4, we have
	\begin{equation}\label{zeta}
		\begin{array}{l}
			{\kern 18pt} {{\bf{W}}_k}{{\bf{r}}_k} = {{\bf{U}}_1}\text{diag}\left( {{{\bf{\Gamma }}_k}} \right){{{\bf{\bar s}}}_k} \vspace{1ex}\\
			\Rightarrow {\bm{\zeta }}_k^*{\bf{r}}_k^H{{\bf{r}}_k} = {{\bf{U}}_1}\text{diag}\left( {{{\bf{\Gamma }}_k}} \right){{{\bf{\bar s}}}_k} \vspace{1ex}\\
			\Rightarrow{\kern 22pt} {\bm{\zeta }}_k^* = \dfrac{1}{{{\bf{r}}_k^H{{\bf{r}}_k}}}{{\bf{U}}_1}\text{diag}\left( {{{\bf{\Gamma }}_k}} \right){{{\bf{\bar s}}}_k}.
		\end{array}
	\end{equation}
	Using (\ref{zeta}), the structure of the optimal receive combining matrix ${{\bf{W}}_k}$ as a function of ${{{\bf{\Gamma }}_k}}$ can be expressed as
	\begin{equation}\label{Wk}
		{{\bf{W}}_k} = {\bm{\zeta }}_k^*{\bf{r}}_k^H = \frac{1}{{{\bf{r}}_k^H{{\bf{r}}_k}}}{{\bf{U}}_1}\text{diag}\left( {{{\bf{\Gamma }}_k}} \right){{\bf{\bar s}}_k}{\bf{r}}_k^H.
	\end{equation}
	With the fact that ${\varphi _k} >0 $ and based on (\ref{c6}), we can obtain that the normalized requirement is strictly active when optimality is achieved. Then, we could have
	\begin{equation}\label{T1}
		\begin{array}{l}
			\left\| {{{\bf{W}}_k}} \right\|_2^2 = 1 \vspace{1ex}\\
			\Rightarrow {\rm{Tr}}\left[ {\dfrac{1}{{{{\left( {{\bf{r}}_k^H{{\bf{r}}_k}} \right)}^2}}}{{\bf{U}}_1}\text{diag}\left( {{{\bf{\Gamma }}_k}} \right){{{\bf{\bar s}}}_k}{\bf{r}}_k^H{{\bf{r}}_k}{\bf{\bar s}}_k^H\text{diag}\left( {{{\bf{\Gamma }}_k}} \right){\bf{U}}_1^H} \right] = 1 \vspace{1ex}\\
			\Rightarrow \dfrac{1}{{{\bf{r}}_k^H{{\bf{r}}_k}}}{\rm{Tr}}\left[ {{{\bf{U}}_1}\text{diag}\left( {{{\bf{\Gamma }}_k}} \right){{{\bf{\bar s}}}_k}{\bf{\bar s}}_k^H\text{diag}\left( {{{\bf{\Gamma }}_k}} \right){\bf{U}}_1^H} \right] = 1 \vspace{1ex}\\
			\Rightarrow \dfrac{1}{{{\bf{r}}_k^H{{\bf{r}}_k}}}{\bf{\bar s}}_k^H\text{diag}\left( {{{\bf{\Gamma }}_k}} \right){\bf{U}}_1^H{{\bf{U}}_1}\text{diag}\left( {{{\bf{\Gamma }}_k}} \right){{{\bf{\bar s}}}_k} = 1 \vspace{1ex}\\
			\Rightarrow \dfrac{1}{{{\bf{r}}_k^H{{\bf{r}}_k}}}{\bf{\Gamma }}_k^T\underbrace {\text{diag}\left( {{\bf{\bar s}}_k^H} \right){\bf{U}}_1^H{{\bf{U}}_1}\text{diag}\left( {{{{\bf{\bar s}}}_k}} \right)}_{{{\bf{T}}_1}}{{\bf{\Gamma }}_k} = 1 \vspace{1ex}\\
			\Rightarrow {\bf{\Gamma }}_k^T{{\bf{T}}_1}{{\bf{\Gamma }}_k} = {\bf{r}}_k^H{{\bf{r}}_k},
		\end{array}
	\end{equation} 
	where ${{\bf{T}}_1} = \text{diag}\left( {{\bf{\bar s}}_k^H} \right){\bf{U}}_1^H{{\bf{U}}_1}\text{diag}\left( {{{{\bf{\bar s}}}_k}} \right)$. Since ${{\bf{T}}_1}$ is a positive semi-definite Hermitian matrix and each entry of ${{\bf{\Gamma }}_k}$ is real, (\ref{T1}) can be further transformed into 
	\begin{equation}
		{\bf{\Gamma }}_k^T{{\bf{T}}_1}{{\bf{\Gamma }}_k} = {\bf{\Gamma }}_k^T{\frak R}\left( {{{\bf{T}}_1}} \right){{\bf{\Gamma }}_k} = {\bf{\Gamma }}_k^T{{\bf{V}}_1}{{\bf{\Gamma }}_k} = {\bf{r}}_k^H{{\bf{r}}_k},
	\end{equation}
	where ${{\bf{V}}_1} = {\frak R}\left( {{{\bf{T}}_1}} \right)$ is a symmetric matrix. Therefore, P4 can be further transformed into an equivalent optimization on ${\bf{\Gamma }}$, given by
	\begin{equation}
		\begin{array}{l}
			\text{P11}:\mathop {\min }\limits_{t,{\bf{\Gamma }}} {\kern 4pt}  - t \vspace{1ex}\\
			{\kern 20pt} s.t.{\kern 10pt} {\bf{C1}}:{\bf{\Gamma }}_k^T{{\bf{V}}_1}{{\bf{\Gamma }}_k} = {\bf{r}}_k^H{{\bf{r}}_k},{\kern 1pt} \forall k \in {\cal K}\vspace{1ex}\\
			{\kern 44pt} {\bf{C2}}:t - \gamma _m^{\cal O} \le 0,\forall \gamma _m^{\cal O} \in {\cal O} \vspace{1ex}\\
			{\kern 44pt} {\bf{C3}}:t - \gamma _n^{\cal I} = 0,\forall \gamma _n^{\cal I} \in {\cal I}.
		\end{array}
	\end{equation}
	
	The optimal receive combining matrix for P4 can be obtained by substituting the solution of P11 into (\ref{Wk}). Additionally, P11 has the same form as ${\mathcal{P}}_4$ in \cite{CF-QAM}, then ${\bf{\Gamma }}_k$ can be further derived based on \textit{Proposition} 2 presented in \cite{CF-QAM}. By substituting the expression for ${\bf{\Gamma }}_k$ into ${{{\bf{W}}_k}}$ in (\ref{Wk}), the closed-form structure (\ref{OptimalWk}) is obtained, which completes the proof.

	\newpage


\begin{thebibliography}{00}
		\balance
		
		\bibitem{Tutorial MIMO}
		H. Lu et al., ``A Tutorial On Near-Field XL-MIMO Communications Towards 6G,'' \textit{IEEE Commun. Surveys Tuts}, Early Access, 2024.
		
		\bibitem{Sensing}
		F. Dong, F. Liu, Y. Cui, W. Wang, K. Han and Z. Wang, ``Sensing as a Service in 6G Perceptive Networks: A Unified Framework for ISAC Resource Allocation,'' \textit{IEEE Trans. Wireless Commun}., vol. 22, no. 5, pp. 3522-3536, May 2023.
		
		\bibitem{Tutorial Large-Scale MIMO}
		Z. Wang et al., ``A Tutorial on Extremely Large-Scale MIMO for 6G: Fundamentals, Signal Processing, and Applications,'' \textit{IEEE Commun. Surveys Tuts}, vol. 26, no. 3, pp. 1560-1605, 2024.
		
		\bibitem{ZF}
		A. Wiesel, Y. C. Eldar, and S. Shamai (Shitz), ``Zero-forcing precoding and generalized inverses,” \textit{IEEE Trans. Signal Process}., vol. 56, no. 9,pp. 4409–4418, Sep. 2008.
		
		\bibitem{RZF}
		C. B. Peel, B. M. Hochwald, and A. L. Swindlehurst, ``A vector perturbation technique for near-capacity multiantenna multiuser communication—Part I: Channel inversion and regularization,” \textit{IEEE Trans. Commun}., vol. 53, no. 1, pp. 195–202, Jan. 2005.
		
		\bibitem{BD}
		Q. H. Spencer, A. L. Swindlehurst, and M. Haardt, ``Zero-forcing methods for downlink spatial multiplexing in multiuser MIMO channels,” \textit{IEEE Trans. Signal Process}., vol. 52, no. 2, pp. 461–471, Feb. 2004.
		
		\bibitem{RBD}
		K. Zu, R. C. de Lamare, and M. Haardt, ``Generalized design of low-complexity block diagonalization type precoding algorithms for multiuser MIMO systems,” \textit{IEEE Trans. Commun}., vol. 61, no. 10,pp. 4232–4242, Oct. 2013.
		
		\bibitem{SLP Surveys1}
		M. Alodeh et al., ``Symbol-Level and Multicast Precoding for Multiuser Multiantenna Downlink: A State-of-the-Art, Classification, and Challenges,'' \textit{IEEE Commun. Surveys Tuts}, vol. 20, no. 3, pp. 1733-1757, 2018.
		
		\bibitem{SLP Surveys}
		A. Li et al., ``A tutorial on interference exploitation via symbol-level precoding: Overview, state-of-the-art and future directions,” \textit{IEEE Commun. Surveys Tuts}., vol. 22, no. 2, pp. 796–839, 3rd Quart., 2020.
		
		\bibitem{Selective-Precoding}
		C. Masouros and E. Alsusa, ``A Novel Transmitter-Based Selective-Precoding Technique for DS/CDMA Systems," \textit{2007 IEEE International Conference on Communications}, Glasgow, UK, pp. 2829-2834, 2007.
		
		\bibitem{DI}
		C. Masouros and E. Alsusa, ``Dynamic linear precoding for the exploitation of known interference in MIMO broadcast systems,” \textit{IEEE Trans. Wireless Commun}., vol. 8, no. 3, pp. 1396–1404, Mar. 2009.
		
		\bibitem{CF-PSK}
		A. Li and C. Masouros, ``Interference exploitation precoding made practical: Optimal closed-form solutions for PSK modulations,” \textit{IEEE Trans. Wireless Commun}., vol. 17, no. 11, pp. 7661–7676, 2018.

		\bibitem{CF-QAM}
		A. Li, C. Masouros, B. Vucetic, Y. Li, and A. L. Swindlehurst, ``Interference exploitation precoding for multi-level modulations: Closed-form solutions,” \textit{IEEE Trans. Commun}., vol. 69, no. 1, pp. 291–308, 2021.
		
		\bibitem{Group SLP}
		Z. Xiao, R. Liu, M. Li, Y. Liu, and Q. Liu, ``Low-complexity designs of symbol-level precoding for MU-MISO systems,'' \textit{IEEE Trans. Commun}.,vol. 70, no. 7, pp. 4624–4639, Jul. 2022.
		
		\bibitem{Large-scale}
		A. Haqiqatnejad, F. Kayhan and B. Ottersten, ``Energy-Efficient Hybrid Symbol-Level Precoding for Large-Scale mmWave Multiuser MIMO Systems,'' \textit{IEEE Trans. Commun}, vol. 69, no. 5, pp. 3119-3134, May 2021.
		
		\bibitem{DACs}
		S. Domouchtsidis, C. G. Tsinos, S. Chatzinotas and B. Ottersten, ``Joint Symbol Level Precoding and Combining for MIMO-OFDM Transceiver Architectures Based on One-Bit DACs and ADCs,'' \textit{IEEE Trans. Commun}, vol. 20, no. 7, pp. 4601-4613, July 2021.
		
		\bibitem{TCOM SER}
		E. S. P. Lopes, L. T. N. Landau and A. Mezghani, ``Minimum Symbol Error Probability Discrete Symbol Level Precoding for MU-MIMO Systems With PSK Modulation,"  \textit{IEEE Trans. Commun}, vol. 71, no. 10, pp. 5935-5949, Oct. 2023.
		
		\bibitem{WCL SER}
		Y. Wang, H. Hou, W. Wang and X. Yi, ``Symbol-Level Precoding for Average SER Minimization in Multiuser MISO Systems," \textit{IEEE Wireless Commun. Lett}., vol. 13, no. 4, pp. 1103-1107, April 2024.
		
		\bibitem{CI Joint Combiner and Precoder}
		Z. Wei, C. Masouros and T. Xu, ``Constructive Interference based Joint Combiner and Precoder Design in Multiuser MIMO Systems,'' \textit{2021 IEEE International Conference on Communications Workshops (ICC Workshops)}, Montreal, QC, Canada, 2021.
		
		\bibitem{PSK SLP MU-MIMO}
		X. Tong, A. Li, L. Lei, F. Liu and F. Dong, ``Symbol-Level Precoding for MU-MIMO System With RIRC Receiver," \textit{IEEE Trans. Commun}, vol. 72, no. 5, pp. 2820-2834, May 2024.
		
		\bibitem{Con JTRSLP}
		S. Cai, T.-H. Chang, and H. Zhu, ``Joint symbol level precoding and receive beamforming for multiuser MIMO downlink,” \textit{Proc. IEEE 20th Int. Workshop Signal Process. Adv. Wireless Commun. (SPAWC)}, Cannes, France, Jul. 2019, pp. 1–5.
		
		\bibitem{Jour JTRSLP}
		S. Cai, H. Zhu, C. Shen, and T.-H. Chang, ``Joint symbol level precoding and receive beamforming optimization for multiuser MIMO downlink,'' \textit{IEEE Trans. Signal Process}., vol. 70, pp. 6185–6199, 2022.
		
		\bibitem{Fifty years mimo detection}
		S. Yang and L. Hanzo, ``Fifty Years of MIMO Detection: The Road to Large-Scale MIMOs," \textit{IEEE Commun. Surveys Tuts}., vol. 17, no. 4, pp. 1941-1988, Fourthquarter 2015.
		
		\bibitem{LOW complxity MIMO Detection}
		K. Kato, K. Fukawa, R. Yamada, H. Suzuki and S. Suyama, ``Low-Complexity MIMO Signal Detection Employing Multistream Constrained Search," \textit{IEEE Trans. Veh. Technol}., vol. 67, no. 2, pp. 1217-1230, Feb 2018.
		
		\bibitem{ZF Detection}
		M. Matthaiou, N. D. Chatzidiamantis, G. K. Karagiannidis and J. A. Nossek, ``ZF Detectors over Correlated K Fading MIMO Channels," \textit{IEEE Trans. Commun}., vol. 59, no. 6, pp. 1591-1603, June 2011.
		
		\bibitem{MMSE Detection}
		Y. Jiang, M. Varanasi, and J. Li, ``Performance analysis of ZF and MMSE equalizers for MIMO systems: An in-depth study of the high SNR regime,” \textit{IEEE Trans. Inf. Theory}., vol. 57, no. 4, pp. 2008–2026, Apr 2011.
		
		\bibitem{MAME Detection}
		R. Lupas and S. Verdú, ``Linear multiuser detectors for synchronous code-division multiple-access channels,” \textit{IEEE Trans. Inf. Theory}., vol. 35, no. 1, pp. 123–136, Jan 1989.
		
		\bibitem{MLD}
		X. Zhu and R. D. Murch, ``Performance analysis of maximum likelihood detection in a MIMO antenna system,” \textit{IEEE Trans. Commun}., vol. 50, no. 2, pp. 187–191, Feb 2002.
		
		\bibitem{QRMLD}
		M. O. Damen, H. E. Gamal, and G. Caire, ``On maximum-likelihood detection and the search for the closest lattice point,” \textit{IEEE Trans. Inf. Theory}., vol. 49, no. 10, pp. 2389–2402, Oct 2003.
		
		\bibitem{QRMMLD}
		 K. J. Kim and J. Yue, ``Joint channel estimation and data detection algorithms for MIMO-OFDM systems,”  \textit{Proc. 36th Asilomar Conf. Signals Syst. Comput}., Pacific Grove, USA, pp. 1857–1861, Nov 2002.
		 
		 \bibitem{Emld}
		 K. Higuchi, H. Kawai, N. Maeda, H. Taoka and M. Sawahashi, ``Experiments on real-time 1-Gb/s packet transmission using MLD-based signal detection in MIMO-OFDM broadband radio access,'' \textit{IEEE J. Sel. Areas Commun}, vol. 24, no. 6, pp. 1141-1153, June 2006.
		 
		 \bibitem{LowComplexityMISO}
		 Z. Xiao, R. Liu, M. Li, Y. Liu and Q. Liu, ``Low-Complexity Designs of Symbol-Level Precoding for MU-MISO Systems,'' \textit{IEEE Trans. Commun}., vol. 70, no. 7, pp. 4624-4639, July 2022.
		 
		 \bibitem{known interference}
		 C. Masouros and G. Zheng, ``Exploiting known interference as green signal power for downlink beamforming optimization,'' \textit{IEEE Trans. Signal Process}., vol. 63, no. 14, pp. 3628–3640, Jul. 2015.
		 
		 \bibitem{interference1}
		 M. Alodeh, S. Chatzinotas and B. Ottersten, ``Symbol-Level Multiuser MISO Precoding for Multi-Level Adaptive Modulation,'' \textit{IEEE Trans. Wireless Commun}., vol. 16, no. 8, pp. 5511-5524, Aug. 2017.
		 
		 \bibitem{MatrixAnalysis}
		 X. Zhang, Matrix analysis and application. Tsinghua, C.H.N.: Tsinghua Univ. Press, 2013.
		 
		 \bibitem{ConvexOptimization}
		 L. Vandenberghe and S. Boyd, Convex Optimization. Cambridge, U.K.: Cambridge Univ. Press, 2004.

		
	\end{thebibliography}
\end{document}